\documentclass[reprint,
tightenlines,
superscriptaddress,
 amsmath,amssymb,aps,pra
]{revtex4-2}
\usepackage{graphicx}
\usepackage{dcolumn}
\usepackage{bm}
\usepackage{natbib}
\usepackage{braket}
\usepackage{amsmath}
\usepackage{amsfonts}
\usepackage{hyperref}
\hypersetup{colorlinks=true,citecolor=blue}
\usepackage{color,xcolor,soul}

\usepackage{orcidlink}
\usepackage{bm}

\usepackage{setspace}

\begin{document}

\title{Vibronic quantum dynamics of ultralong-range high-$\ell$ Rydberg molecules}

\author{Felix Giering}
\affiliation{Zentrum für Optische Quantentechnologien, Universität Hamburg, Luruper Chaussee 149, 22761 Hamburg, Germany}

\author{Rohan Srikumar\orcidlink{0000-0003-0303-1331}}
\affiliation{Zentrum für Optische Quantentechnologien, Universität Hamburg, Luruper Chaussee 149, 22761 Hamburg, Germany}

\author{Peter Schmelcher\orcidlink{0000-0002-2637-0937}}
\affiliation{Zentrum für Optische Quantentechnologien, Universität Hamburg, Luruper Chaussee 149, 22761 Hamburg, Germany}

\date{\today}

\begin{abstract}

We investigate the non-adiabatic quantum dynamics of ultralong-range Rydberg molecules using a vibronically coupled two-channel treatment. The two-channels are composed of coupled trilobite and butterfly electronic states, formed as a result of $S$-wave and $P$-wave scattering of high angular momentum Rydberg electrons with perturbing ground state atoms. Within the Born-Oppenheimer treatment, the $P$-wave scattering channel introduces an adiabatic decay pathway that affects the stability and lifetimes of trilobite states. Our numerical results show that the vibronic coupling is dependent on the principal quantum number $n$, and for certain $n$ there is non-adiabatic stabilization against internal molecular decay, facilitating previously studied dynamical effects in pure trilobite molecules. Apart from the internal diffraction effect caused by the backscattering of vibrational wavepackets from the trilobite potential, we also observe interesting multi-well tunneling effects during low-energy oscillations for a range of $n$-values. Our work serves to highlight that the unique vibrational coordinate dependent electronic structure of these polar molecules, along with high level densities, promise many exciting dynamical effects.

\end{abstract}

\maketitle

\section{Introduction}

Long-range synthetic molecules formed by
engineering Rydberg excitations in ultracold atomic gases
present an exciting frontier in few-body quantum physics research
\cite{Greene2000,Hamilton_2002,Hollerith_2021,hollerith_2019,anasuri_2023,Zuber2022,bosworth_2022,Rosario2021,Wang2015,Stanojevic2006,Stanojevic2008}, 
exhibiting properties not associated with traditional molecular systems. Of recent fame are the exemplar trilobite and butterfly molecules, formed when a Rydberg atom with high electronic angular momentum ($\ell$) undergoes $S$-wave and $P$-wave scattering from a nearby ground state atom (perturber), respectively \cite{Greene2000,Hamilton_2002}. They feature exceptionally high dipole moments \cite{Booth_2015,Sadeghpour_2011} and micrometer scale bond-lengths and have been found useful in precision spectroscopy\cite{Exner2024,Engel_2019}, prospective preparation of ultracold neutral plasmas \cite{Hummel_2020}, and measuring three-body correlations \cite{Killian2023,Fey2019}. Moreover, these ultralong-range Rydberg molecules (ULRMs) \cite{Hummel2020,Shaffer2018,Eiles_2019} offer a great testbed to study unique and exaggerated quantum dynamics \cite{Keiler_2021,Fey2016,srikumar2024}, and external field interactions \cite{Lesanovsky_2006,Niederprum2016,Kurz_2013}. Recent numerical simulations show that nuclear wavepackets propagating through the undulating trilobite potential energy curves (PECs) undergo back scattering, thereby creating an internal diffraction effect wherein the electronic structure of the dimer diffracts its own radial motion \cite{srikumar2024}, a phenomena unobserved in standard molecular physics.  \\

\begin{figure}
    \centering
    \includegraphics[width=\linewidth]{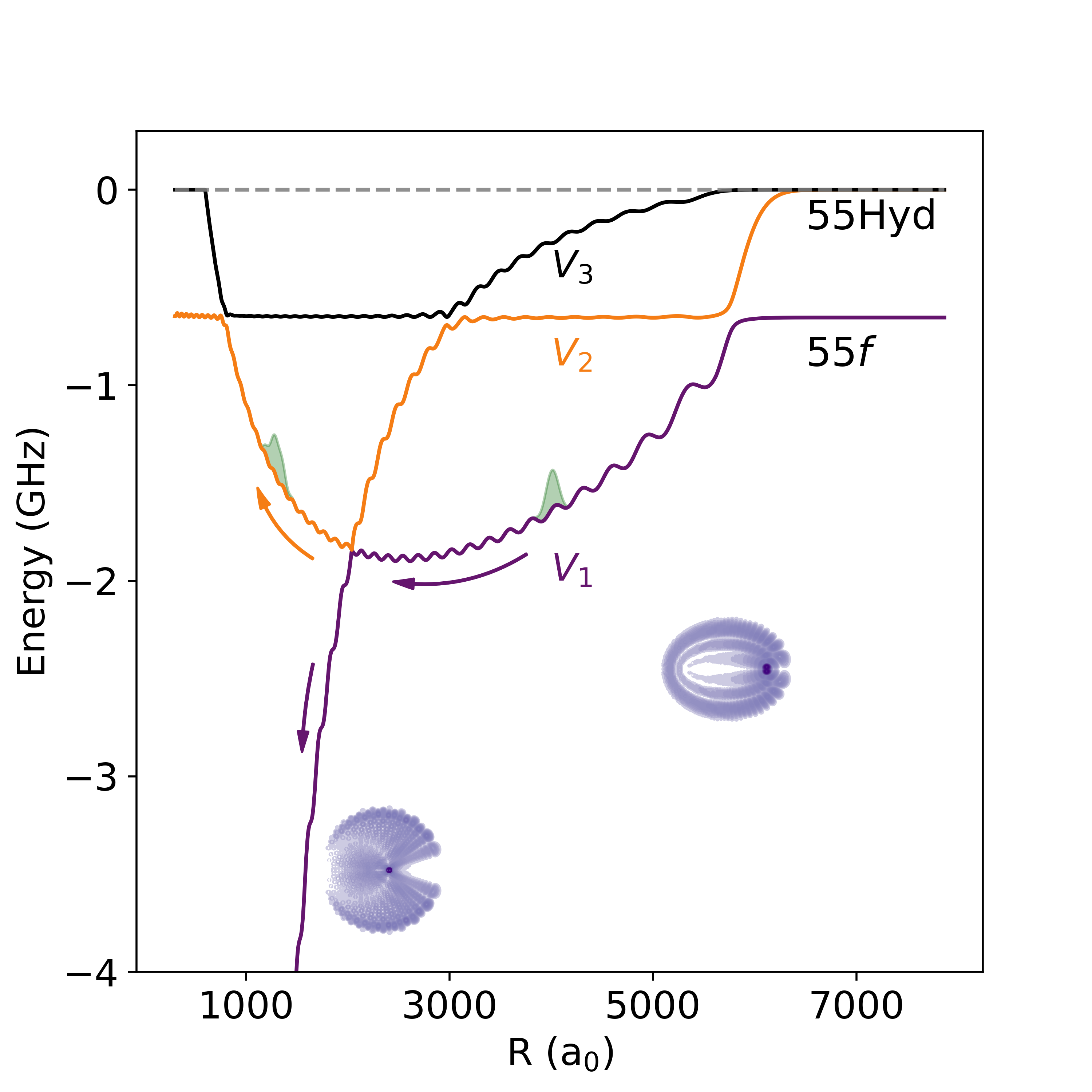}
    \caption{The Born-Oppenheimer potentials of the $n$=55 ULRM, obtained by solving the electronic Schrödinger equation. The adiabatic potentials are highlighted as $V_1$ (violet), $V_2$ (orange) , and $V_3$ (black), whereas the grey dashed curves show the unperturbed hydrogenic manifold. The wavepackets and arrows are used to illustrate the possible trajectories of a time evolving nuclear wavefunction. The probability density plots of $\psi_1(\textbf{r},R)$ (corresponding to the PEC $V_1$) calculated at $R$=1400 a$_0$ and $R$=4880 a$_0$, portray the change of electronic structure from trilobite to butterfly states as $R$ changes. }
    \label{fig:potential_energy}
\end{figure}

The traditional approach to study the quantum dynamics of ULRM is through the use of the Born-Oppenheimer (BO-) approximation \cite{Born_1927}. However, recent studies show that non-adiabatic effects are ubiquitous in such systems, mainly due to avoided crossings introduced by high-$\ell$ states, as well as the high density of electronic states that are not energetically well-separated \cite{Hummel_2022,Srikumar2023}. Such avoided crossings can be highly singular, sometimes forming conical intersections, when the principal quantum number $n$ of the Rydberg atom is taken as a synthetic dimension \cite{Hummel2021,eiles2024}.  
In alkali ULRMs, the $P$-wave butterfly potential induces an avoided crossing to the trilobite and low$-\ell$ potentials, introducing an adiabatic decay pathway. This affects the lifetimes of bound ULRM states \cite{Butscher_2011}, while also allowing for internal quantum reflection \cite{Bendkowsky2010}. Recent experimental observations of highly excited non-adiabatic resonances stabilized against internal adiabatic decay \cite{srikumar2025},
confirm the necessity to go beyond single channel BO-approximation to accurately model these molecules.  
Apart from atom-atom ULRM, there is also an increasing interest in the study of non-adiabatic phenomena arising in other long-range Rydberg systems, like Rydberg atom-ion molecules \cite{Raithel2022}, macrodimers \cite{hollerith_2019,Samboy2011}, Rydberg bimolecules \cite{Rosario2024}, and trapped Rydberg ion systems \cite{Gambetta2021,Wilkinson2024}. \\

The multichannel wavepacket dynamics in ULRM remains largely unexplored. Previous works have mainly utilized Landau-Zener based semiclassical analysis \cite{Pfau2016,Engel2024,Hummel2021}, or limited themselves to spectral calculations \cite{Hummel_2022,Srikumar2023,srikumar2025,durst2024}. An exact quantum dynamical study will be able to validate previous semiclassical approximations, as well as point out why and where they fail. Moreover, it confirms whether or not previously simulated single channel dynamical effects, like the internal diffraction of vibrational wavepackets in the trilobite potential \cite{srikumar2024}, are suppressed due to non-adiabatic effects. In conventional molecules vibronic dynamics facilitate many well documented phenomena like ultrafast non-radiative transitions \cite{Marciniak2015,Koppel1983,Domcke1993}, surface hoppings \cite{Herman_1984,Barbatti_2011}, and pre-dissociation \cite{Rice1933,Harris1963}, which are not explainable by adiabatic physics. In atom-ion scattering, vibronic dynamics is also necessarily included in studies of charge exchange collisions \cite{Peng2011}.
But highly polar ULRMs, with large bond lengths and high level densities could exhibit new vibronic dynamical effects that are not seen in regular molecules or scattering processes. \\

In this work we study the dynamics of $^{87}$Rb dimers, in coupled  trilobite and butterfly configurations using a two-channel treatment that considers the mixing of nuclear and electronic motion. A comprehensive theoretical study of this system, which looks at the structure of electronic states and non-adiabatic couplings as a function of $n$, develops diabatization strategies to work around singular couplings, and calculates the molecular vibronic spectra  has already been presented \cite{Hummel_2022}. Here we perform wave-packet dynamics, with vibrational Gaussian states initialized in the trilobite PEC, radially far off from any avoided crossings. As the wavepackets probe the avoided crossing, we see the onset of $n$-dependent non-adiabatic transitions between different PECs.
We use the probability distribution function in position and momentum representations, as well as the evolution of different electronic state populations, to characterize vibronic effects for each $n$. For some $n$, the propagating diffraction effect is conserved against internal molecular decay due to non-adiabatic stabilization. We also present here a unique multi-well tunneling phenomenon due to low-energy oscillations around the minima of the trilobite potential. For a range of $n$-values, these trilobite curves exhibit bound vibrational states localized between multiple wells, near the potential minima; allowing unique interference effects in their low-energy dynamics. In comparison trilobite PECs with much smaller $n$-values  ($\leq$ 30) primarily exhibit low-energy dynamics confined to a single well. The calculations presented here neglect the fine and hyperfine structure of the constituent atoms for the sake of ease. Inclusion of the spin couplings would necessitate a more complex multichannel treatment,  relevant to future studies.\\

\begin{figure}
    \centering
    \includegraphics[width=\linewidth]{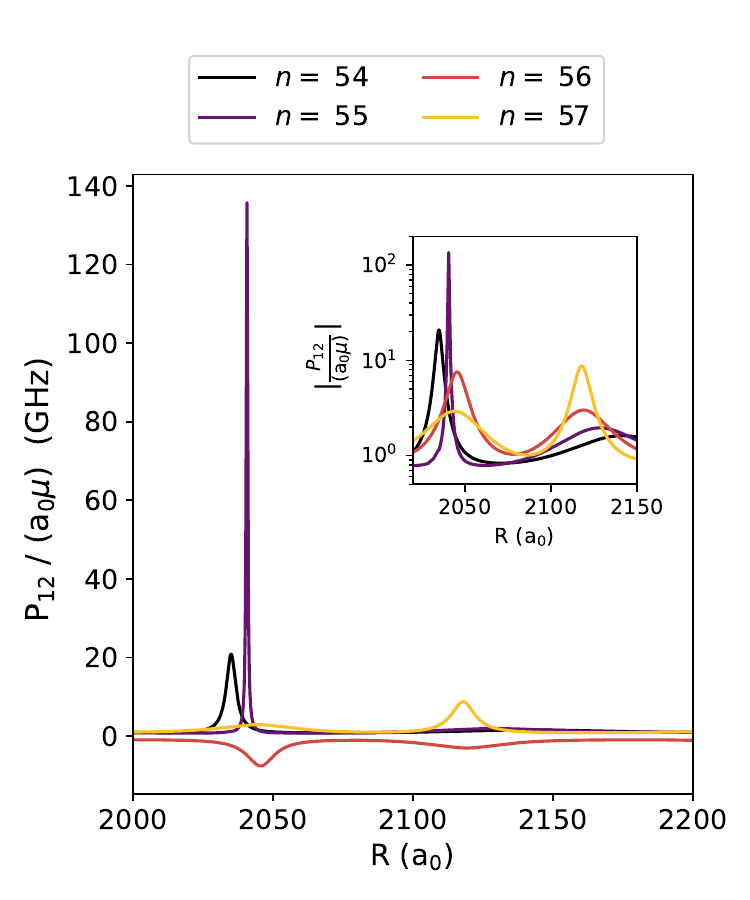}
    \caption{ The first order derivative coupling element $P_{12}/(\mu \mathrm{a}_0)$ as a function of the internuclear) distance $R$, for varying $n$-values. The inset depicts the magnitude of $P_{12}$ with log-scaling of the $y$-axis. }
    \label{fig:non_adiabatic}
\end{figure}

This work is organised as follows. Section \ref{Sec2} presents the theoretical and computational framework, introducing the general molecular Hamiltonian and nonadiabatic couplings in subsection \ref{Sec2A}, discussing the specific electronic interactions in the ULRM in subsection \ref{Sec2B}, and describing the two-channel vibronic dynamics in subsection \ref{Sec2C}. Section \ref{Sec3} presents the results of our numerical simulation, starting from the potential energy structure in subsection \ref{Sec3A}, the two-channel wavepacket dynamics in subsection \ref{Sec3B}, and multi-well tunnelling effects in subsection \ref{Sec3C}. The conclusions and outlook of our work are given in section \ref{Sec4}.

\section{Theoretical and computational framework} \label{Sec2}

We begin by formulating the molecular Hamiltonian and the corresponding time dependent Schrödinger equation which allows us to study the internal molecular dynamics of ULRM. We then discuss the Born-Oppenheimer approximation, the emergence of nonadiabatic couplings between BO adiabatic electronic states, and the associated diabatization procedure which is crucial for constructing a smooth multi-channel representation. Subsequently, we highlight the electronic structure of the ULRM and the calculation of adiabatic states and PECs. Finally, we establish the two-channel vibronic system and describe the numerical propagation scheme, where complex absorbing potentials are employed to avoid artificial reflections at the grid boundaries. 

\subsection{Molecular Hamiltonian and nonadiabatic couplings} \label{Sec2A}

We start by considering two $^{87}$Rb atoms, one of which is excited to the Rydberg electronic state. Let us define $R$ as the internuclear separation between the ionic core and the neutral atom (taken along the z-axis, with the Rydberg core at the origin), and $\textbf{r}$ as the coordinate of the Rydberg electron with respect to the Rydberg core. The internal molecular dynamics (in atomic units) in the body-fixed frame is then governed by the time-dependent Schr\"odinger equation
\begin{equation}
    i \frac{\partial}{\partial t} \Psi_m (\bm{r};R \bm{e}_z, t) 
    = \hat{H} \Psi_m (\bm{r};R \bm{e}_z, t) \, ,
\end{equation}
with $\Psi_m$ being the molecular wavefunction and $\hat{H}$ the molecular Hamiltonian given by:
\begin{equation} \label{eq:molecular_hamiltonian}
    \hat{H} = - \frac{\nabla^2_R}{2\mu} - \frac{\nabla^2_{\bm{r}}}{2} + V_\mathrm{Ryd}(\bm{r}) + V_\mathrm{ea}(\bm{r};R \bm{e}_z)
\end{equation}
where $\mu$ is the reduced mass of the system. The first two terms of the Hamiltonian describe the kinetic energy operators corresponding to the nuclei and the Rydberg electron respectively. Moreover, $V_\mathrm{Ryd}(r)$ represents the interaction between the Rydberg electron and the ionic core and $V_\mathrm{ea}(\textbf{r};R)$ accounts for the interaction between the Rydberg electron and the perturbing ground state atom. The last three terms of Eq.~\ref{eq:molecular_hamiltonian} separately form the electronic Hamiltonian $H_\mathrm{el} (\textbf{r};R)$. Solving the electronic Schrödinger equation parametrized by $R$ yields
\begin{equation}
    H_\mathrm{el}(\bm{r};R)\,\psi_i(\bm{r};R) = V_i(R)\,\psi_i(\bm{r};R),
\end{equation}
where $V_i(R)$ are the adiabatic potential energy curves and $\psi_i(\bm{r};R)$ are the corresponding electronic states.
Since $\{ \psi_i\}$ forms a complete orthonormal basis (commonly known as the Born-Oppenheimer basis), any given molecular wavefunction can be expanded as $\Psi_{m} = \sum_{i} \chi_{n}(R) \psi_{i}(\bm{r},R) $ \cite{Huang_1954}. Here $\chi_{i}(R)$ are the $R$-dependent expansion coefficients that represent the nuclear wavefunction.
\newline

\begin{figure*}
    \centering
    \includegraphics[width=0.8\linewidth]{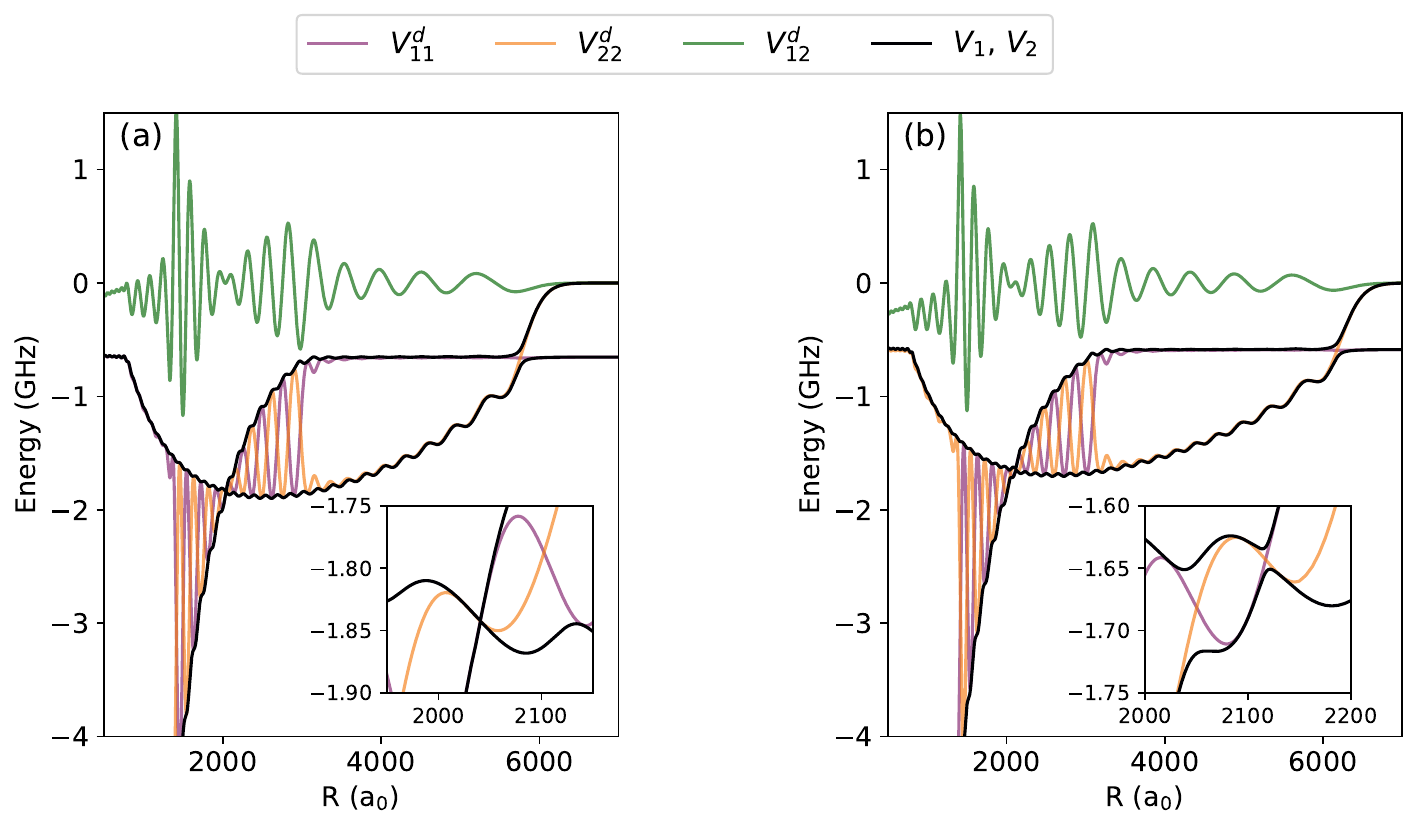}
    \caption{The adiabatic and diabatic PECs of the (a) $n=55$ and (b)  $n=57$ ULRM are shown, with the corresponding insets showcasing a magnified plot of the avoided crossing. The black curves are the adiabatic potentials ($V_1, V_2$) and the colored curves are the diagonal ($V^d_{11}, V^d_{22}$) and off-diagonal ($V^d_{12}$) terms of the diabatic potential energy operator. }
    \label{fig:diabatic}
\end{figure*}

Inserting this Born-Oppenheimer expansion in Eq.~\ref{eq:molecular_hamiltonian},  integrating out the electronic states and rearranging the terms, we get the multichannel Schrödinger equation
\begin{equation} \label{eq:multichannel_hamiltonian}
     - \frac{1}{2\mu}(\nabla_R + \hat{P})^2 \bm{\chi} + \hat{V} \bm{\chi} = i \frac{\partial}{\partial t} \bm{\chi}
\end{equation}
where $\bm{\chi}$ is a vector of all $\chi_{n}$, $\hat{V}$ is a diagonal operator containing all $V_n$ and $\hat{P}$ is the first order non-adiabatic coupling operator whose matrix elements are given by 
\begin{equation} \label{eq:Pmatrix}
    P_{ij}(R) = \int \psi_i(\bm{r};R)  \partial_R \psi_j(\bm{r};R)  d\bm{r} \,.
\end{equation}
The $P$-matrix elements appear as off-diagonal terms in the nuclear kinetic energy operator, effectively coupling different adiabatic electronic channels.
Note that a more elaborate discussion on derivative couplings can be found in our previous works \cite{Srikumar2023,Hummel_2022}, and methodical derivations that show how to obtain Eq.~\ref{eq:multichannel_hamiltonian} from Eq.~\ref{eq:molecular_hamiltonian} are discussed in standard textbooks \cite{Baer_2006,Domcke2004}. In the Born-Oppenheimer approximation, the non-adiabatic term $\hat{P}$ is completely ignored, resulting in decoupled single channel equations that describe the nuclear dynamics of the molecule.  This framework provides a natural first description of ultralong-range Rydberg molecules, where the diatomic system time-evolves in a given potential energy curve defined by its time-independent electronic state.

However, non-adiabatic couplings become particularly pronounced near avoided crossings of the adiabatic PECs. Our previous works highlight such cases where a non-adiabatic treatment of the ULRM was necessary due to the presence of avoided crossings \cite{Srikumar2023,Hummel_2022,Hummel_2019}. In such regions the BO approximation breaks down due to singular or rapidly varying couplings, and the adiabatic basis becomes ill-suited for describing nuclear dynamics. To remedy this, it is advantageous to perform a basis transformation such that the nuclear kinetic couplings are replaced with off-diagonal potential terms. Given that  Eq.~\ref{eq:multichannel_hamiltonian} is invariant under unitary transformations, any $\hat{U}_d$ which satisfies the condition $\nabla \hat{U}^d = - \hat{P} \hat{U}^d$ will force the derivative couplings to become zero, and is commonly known as a diabatic transform \cite{koppel_1984,Baer_2006}. This allows us to rewrite Eq.~\ref{eq:multichannel_hamiltonian} as
\begin{equation} \label{eq:diabatic_hamiltonian}
     - \frac{1}{2\mu}\nabla_R^2 \bm{\chi}^{d} + \hat{V}^{d} \bm{\chi}^{d} = i \frac{\partial}{\partial t}\bm{\chi}^{d} \,,
\end{equation}
where $\bm{\chi}^{d}  =  \hat{U}^d \bm{\chi}$ are called the diabatic states and the constituents of $ \hat{V}^{d} =  \hat{U}^{\dagger}_d \hat{V}  \hat{U}_d$ are called diabatic PECs. In the general case of polyatomic molecules, $\hat{U}^d$ is not uniquely defined and approximate path-dependent diabatization schemes are used to perform vibronic dynamics. Fortunately, if the system is one-dimensional (like a diatomic molecule) one can always find a unique diabatization scheme numerically, although this becomes computationally demanding with increasing number of highly coupled channels. Furthermore, in the simplest description of a vibronic system, i.e.~a two-channel system with a single parameter $R$, the diabatic transform is a basic unitary rotation where the rotation angle is given by
\begin{equation}
\theta(R) = \int_{R}^{\infty} {P_{12}(R_1) dR_1} \,.
\end{equation}

With the two-channel diabatic framework established, the next step is to specify the electronic interactions responsible for binding in ultralong-range Rydberg molecules.

\subsection{Electronic interaction} \label{Sec2B}

\begin{figure*}
    \centering
    \includegraphics[width=0.8\linewidth]{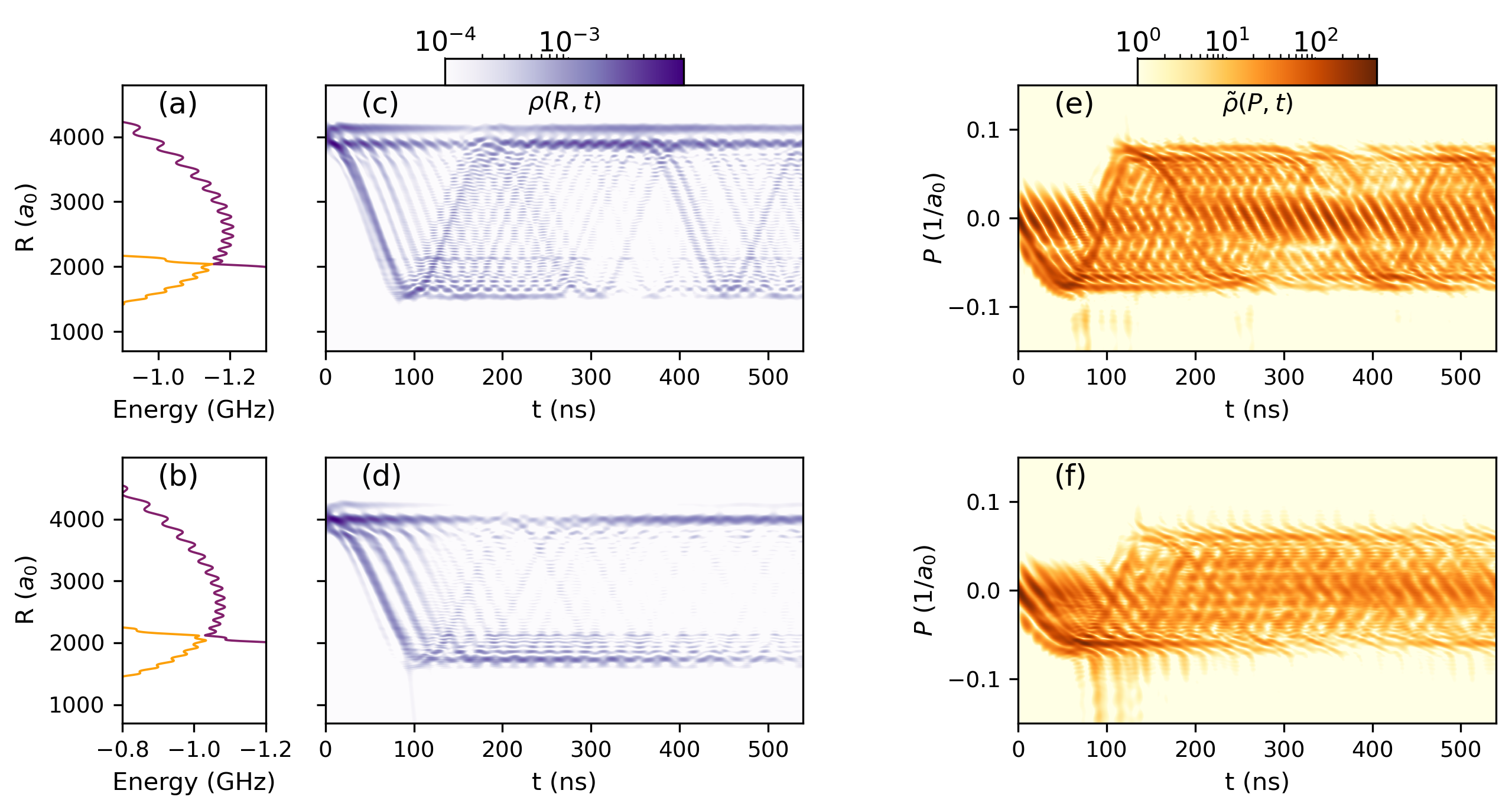}
    \caption{Two-channel wavepacket diffraction dynamics. The potential energy structure of the $n=55$ and $n=57$ ULRM in the relevant region, are shown in (a) and (b). The  time evolving total probability density $\rho(R,t)$ in the position representation is shown for $n=55$ (c) and $n=57$ (d). The corresponding total probability density in momentum representation, i.e.~$\tilde{\rho}(P,t)$ is shown for $n=55$ (e) and $n=57$ (f).   }
    \label{fig:probability_density}
\end{figure*}

The electronic Hamiltonian for the ULRM is given by $H_\mathrm{el} = H_\mathrm{Ryd} + V_\mathrm{ea}$, where $H_\mathrm{Ryd}$ describes the Rydberg atom and $V_\mathrm{ea}$ describes the low energy electron-atom scattering interaction. The $S$-wave and $P$-wave scattering can be modeled using Fermi-type pseudo-potentials \cite{Fermi1934,Omont_1977} such that

\begin{align}
    V_\mathrm{ea}(\bm{r};R)\, &= \,2 \pi a_S(k) \delta(\bm{r}-R \bm{e}_z) \notag \\ & +  \, 6 \pi a_P(k) \nabla_r   \delta(\bm{r} - R \bm{e}_z) \nabla_r \, ,
\end{align}
with $a_S$ and $a_P$ denoting the energy dependent $S$-wave scattering length and $P$-wave scattering volume respectively (scattering length taken from \cite{Engel_2019}). The electron wavenumber $k$ can be semi-classically approximated by $k = ({2}/{R} - {1}/{n^2})^{1/2}$, where $n$ is the principal quantum number of the Rydberg excitation. In the spin triplet configuration, both $a_S$ and $a_P$ have negative values, facilitating an attractive interaction and binding the molecule. 
The unperturbed Rydberg electron states $\{ \phi_{n\ell m}\}$ are described by the principal and angular momentum quantum numbers $(n,\ell,m)$ respectively. Since the internuclear axis is taken to be along the z-axis, only atomic states with $m$=0 contribute. They form the eigenbasis of the electronic Rydberg Hamiltonian with the corresponding energies $E_{n\ell} = - (n - \delta_{\ell})^{-2}/2$,
with $\delta_{\ell}$ denoting the $\ell$-dependent quantum defect \cite{Gallagher2003}. In rubidium the low angular momentum states ($\ell \leq 3$) exhibit finite quantum defects, energetically shifting them  away from the near-degenerate hydrogenic manifold of high-$\ell$ states. These low-$\ell$ states form shallow PECs allowing a few bound states. The high-$\ell$ states however is hybridized by the perturber, resulting in deep potentials that can hold hundreds of bound states. The most prominent of these high-$\ell$ states are the so-called trilobite and butterfly molecules, split from the hydrogenic manifold due to $S$-wave and $P$-wave electron-atom scattering. \\

Recent works have already characterized the vibronic structure of the coupled trilobite-butterfly system,  successfully showing the impact of these avoided crossings in the calculated and observed spectra \cite{Hummel_2022,srikumar2025}.
If one is to ignore the quantum defect structure and limit the system to a single  hydrogenic $n$-manifold, one can analytically derive a coupled two-channel model based on the pure trilobite and butterfly wavefunctions
\begin{align}
    \psi_t (\bm{r};R) &= \frac{1}{\sqrt{\sum_{\ell} |\phi_{n\ell 0} (R)|^2}}\sum_{\ell}^{n-1} 
    \phi_{n\ell 0}(\bm{r})  \phi_{n\ell 0} (R) \\
    \psi_b (\bm{r};R) &= \frac{1}{\sqrt{\sum_{\ell} |\partial_R\phi_{n\ell 0} (R)|^2}}\sum_{\ell} ^{n-1}
    \phi_{n\ell 0}(\bm{r}) \partial_R \phi_{n\ell 0} (R) \, ,
\end{align}
respectively. However, since the above given states are not orthogonal to each other, expanding the electronic Hamiltonian  in the $\{\ket{\psi_t},\ket{\psi_b} \}$ basis is not straightforward, and does not guarantee the validity of the formalism presented in subsection \ref{Sec2A}. Fortunately, since $V_\mathrm{ea}$ is a rank-2 operator, its image space can be spanned by any constituent pair of linearly independent wavefunctions. Recently, it was shown that \cite{eiles2024} one can utilize a semi-unitary transform to project the matrix representation of $V_\mathrm{ea}$ in the hydrogen atom basis to a perturber spherical basis, following which one obtains the potential energy operator 
\begin{align}
\label{eq:hmat}
    V = \begin{pmatrix}
        V_t & V_c \\ V_c & V_b
    \end{pmatrix} \, ,
\end{align}
where $V_t = 2\pi a_S(k)\sum|\phi_{n\ell0}(R)|^2 $ and 
$V_b = 6\pi a_P(k)\sum|\partial_R\phi_{n\ell0}(R)|^2 $ are the trilobite and butterfly PECs, and $V_c =\sqrt{12\pi^2a_S(k)a_P^3(k)}\sum\phi_{n\ell0}(R) \partial_R \phi_{n\ell 0}(R)$ is the coupling between them \cite{eiles2024,srikumar2025}. We will obtain the two channel adiabatic PECs and electronic states by diagonalizing the above given matrix. However, it is already clear that the coupled trilobite-butterfly system exhibits conical intersections at $(n,R$) coordinates, when $V_t = V_b$ and $V_c = 0$ simultaneously.   The first condition is roughly fulfilled when the $S$-wave and $P$-wave scattering phases are equal \cite{Borodin_1992}, and the second condition is fulfilled when the hydrogenic $ns$-state has a node at $R$ (derived using Coulomb summation rules \cite{Ermolaev2000,Cherkani_2001}). 
This model was used to successfully predict the $n$-dependent behavior of the avoided crossing width and the derivative coupling element, which in turn affected the experimentally observed stability of non-adiabatic resonance states  \cite{srikumar2025}. However, in view of performing exact quantum dynamics, we do not utilize this hydrogenic approximation. Our numerical model considers the quantum defect structure of low-$\ell$ states, and is not limited to a single $n$-manifold. The electronic spectra and non-adiabatic couplings exhibited by such a model has also been discussed comprehensively in Ref.~\cite{Hummel_2022}.
This work is exclusively focused on extending these studies by performing vibrational dynamics simulations in the coupled butterfly-trilobite configuration.
\newline

Here we mention that the state of the art Green’s function methods \cite{eiles2023} provide a more accurate description of the electronic structure of the ULRM, however they do not allow for the calculation of the electronic states and their couplings so far. Moreover, spin–orbit and hyperfine interactions \cite{Eiles_2017,Anderson2014} are neglected in our model so that a two-channel treatment becomes possible. Fortunately, recent works show that a simple spin-less  two-channel system can already explain observed non-adiabatic phenomena in ULRM very well \cite{srikumar2025}.
Hence, we choose to focus only on the two-channel case relevant for the trilobite–butterfly interaction, which is well-justified when the dynamics is dominated by a pair of energetically isolated states with localized coupling, while all other channels remain weakly interacting and well-separated. As a first study we choose to focus on a two-channel model that includes the primary non-adiabatic correction to the dynamics, introduced by the avoided crossing between trilobite and butterfly states. For comparison, the non-adiabatic couplings obtained using a three channel model are discussed in the Appendix \ref{A2}.

\subsection{Computational approach: two-channel vibronic dynamics} \label{Sec2C}

We start by obtaining the adiabatic PECs and the corresponding electronic wave functions for a given $n$ by diagonalizing the electronic Hamiltonian within a finite Rydberg basis $\{\phi_{n'\ell ' 0}\}$ that includes all states with $n' \in \{n-2,n+2\} $. Then we proceed to calculate the diabatic transform and the diabatic PECs for the two-channel trilobite-butterfly system.  The initial nuclear wavepacket is a Gaussian function in the first adiabatic channel, i.e.~ $\bm{\chi^a} = (\chi^{a}_1, \,\chi^{a}_2)^T$ with

\begin{align}
    \chi^{a}_1(R,0)& = 
    \frac{1}{(\pi \sigma^2)^{1/4}}
    \exp\!\left[-\frac{(R-R_0)^2}{2\sigma^2} \right], \\
    \chi^{a}_2(R,0)&=0 \, . \label{Eq:vibronic_chi}
\end{align}

Initializing the wavepacket in the adiabatic basis allows us to populate the trilobite or butterfly state selectively, which is made difficult in the diabatic basis due to strong mixing of electronic states.
The adiabatic wavefunction is then transformed into the diabatic basis via $\bm{\chi}^{d}(R,t=0) = \hat{U}(R)\, \boldsymbol{\chi}^{a}(R,t=0) \,$. We then propagate  $\bm{\chi}^{d}(R,t)$  using the diabatic Schrödinger equation given by Eq.~\ref{eq:diabatic_hamiltonian}.
\newline

Time propagation of the wavepacket is performed using the Crank–Nicolson scheme \cite{Koonin1990,thomas2013numerical}, which provides stable and unitary time evolution. The second derivative is evaluated using finite differences of tenth order \cite{Groenenboom_1990}, ensuring high numerical accuracy even for small grid spacings. In our simulations, we use a grid spacing of $\Delta R \approx 1$a$_0$ and a time step of $\Delta t \approx 0.001 ~\text{nanoseconds}$, chosen such that the energy scale of the system is well resolved \cite{Feit1982}.
Quantum reflection at the grid boundaries can lead to unphysical interference in the wavepacket dynamics. To prevent this, we employ a complex absorbing potential (CAP) \cite{mctdh_intro, Riss1993} that smoothly absorbs the wavefunction near the grid edge and thus simulates the molecular decay due to the open channel nature of our system. See Appendix \ref{A1} for more details on the CAP implementation.
\\

At every time step the total probability density is calculated to be
\begin{equation}
    \rho(R,t) = |\chi^d_1(R,t)|^2 + |\chi^d_2(R,t)|^2 \, ,
\end{equation}
and the diabatic channel populations are calculated to be
\begin{equation}
p^d_i(t) = \int \! dR \, |\chi_i^{d}(R,t)|^2 \,.
\label{eq:pop_dia}
\end{equation} 
One can perform an inverse diabatization transform $\bm{\chi}^{a} = U^\dagger\, \bm{\chi}^{d} \,$ to obtain the adiabatic channel populations
\begin{equation}
 p^a_i(t) = \int \! dR \, |\chi_i^{a}(R,t)|^2 \, 
 \label{eq:pop_adia}
\end{equation}
which help us understand how the electronic states are mixed as a function of time due to vibronic dynamics. 
Due to the CAP the time evolution is no longer unitary and the total population
\begin{equation}
    p_{tot}(t) = \sum_i p^a_i(t) = \sum_i p^d_i(t)
     \label{eq:pop_tot}
\end{equation}
does not maintain unit value, as the wavefunction is absorbed at the end of the grid.
Note that the total probability density  $\rho(R)$ as well as the total population $p_{tot}$ are basis independent quantities and hence invariant under the diabatization transform. To understand the scattering effects in the two-channel problem, one can use the Fourier transform to calculate the momentum space wavefunction $\tilde{\chi}(P,t) = \mathcal{F} [\chi (R,t)]$ in each channel, and consequently obtain the total momentum space probability distribution $\tilde{\rho}(P,t) = |\tilde{\chi_1}(P,t)|^2 + |\tilde{\chi_2}(P,t)|^2$, where $P$ is the momentum of the nuclear wavepacket. In the next section, we numerically investigate the vibronic dynamics of the ULRM within the context of internal diffraction dynamics, multi-well tunnelling effects, and non-adiabatic decay.

\section{Results and discussions} \label{Sec3}

In the following section, we discuss the results of our numerical simulations. We start with the adiabatic and diabatic potential energy curves as well as the non-adiabatic couplings in the coupled trilobite-butterfly system. We then study the multichannel internal diffraction effect, and analyze the $n$-dependent non-adiabatic corrections by studying the probability density functions of the evolving wavepacket as well as the population dynamics. Finally, we unravel multi-well tunneling effects in the low energy oscillations around the minima of the trilobite potential.

\subsection{Potential energy structure} \label{Sec3A}

We start with a brief review of the electronic structure of high-$\ell$ ULRM. Figure \ref{fig:potential_energy}  shows the adiabatic PECs for an $n=55$ ULRM obtained by diagonalizing the electronic Hamiltonian, with the exemplary $n$-value being ideal for observing internal diffraction effects. The adiabatic PECs relevant to the two-state model are highlighted in color ($V_1$ as blue and $V_2$ as red), whereas states corresponding to the grey adiabatic curves are not considered in further calculations. When the perturber is radially outside the Rydberg atomic orbital, ($R \geq 2n^2$), the adiabatic potentials $V_1$ and $V_2$ energetically coincide with the unperturbed $55f$ quantum defect state and the high-$\ell$ hydrogenic state, respectively. At smaller $R$-values the $S$-wave  scattering hybridizes the hydrogenic manifold, introducing an avoided crossing between the trilobite PEC and the $\ell =3$ PEC, forming the $V_1$ potential. Here, we choose to focus only on the two highlighted curves as a first approximation. Without the $P$-wave interaction, $V_1$ would be a pure trilobite curve with the structure of an anharmonic envelop overlayed with an oscillatory potential \cite{srikumar2024}. However, the $P$-wave shape resonance induces a steeply descending butterfly potential that forms a narrow avoided crossing between $V_1$ and $V_2$. After the avoided crossing $V_1$ adiabatically acquires the $P$-wave character, eventually causing internal molecular decay due to autoionization or $\ell$-changing collisions \cite{Pfau2016}. However, we know that a narrow avoided crossing enhances the vibronic coupling terms, enabling the non-adiabatic transition from $V_1$ to $V_2$, possibly offering a stabilizing mechanism against internal molecular decay. The adiabatic electronic probability density plots of $\psi_1(\textbf{r},R) $ calculated sufficiently before and after the avoided crossing, show how $\psi_1$ adiabatically changes from trilobite to butterfly configuration. \newline

Figure \ref{fig:non_adiabatic} shows the non-adiabatic coupling $P_{12}$ between the selected states $\psi_1$, $\psi_2 $ for the selected range of $n$. We scale $P_{12}$ by $1/(\mu \mathrm{a}_0)$ to analyze the energy contribution of the non-adiabatic coupling term as it appears in the Hamiltonian, in GHz units. We see that the derivative coupling strongly depends on $n$, displaying an order of magnitude difference between the most non-adiabatic ($n=55$) and most adiabatic($n=57$) cases. This is highlighted in the inset, where the $y$-axis is log-scaled to simultaneously show the smaller structures in energy. From a semi-classical perspective, the absolute value of the $P_{12}$, at the avoided cross massively impacts the probability that nuclei perform a non-adiabatic jump, i.e.~Landau-Zener (LZ-) transition from $\psi_1$ to $\psi_2$. However, we already see from the inset that the calculated derivative couplings do not form a Lorentzian curve, as is typical in LZ-transitions.
This is because the level spacing and the derivative coupling elements are comprised of hybridized atomic wavefuctions and their derivatives \cite{Hummel_2022}, creating an oscillatory non-Lorentzian structure. The standard Landau-Zener theory on the other hand assumes a linear crossing with a well defined constant velocity. This is clearly not what happens in low-energy dynamics of ULRM, where the time evolving wavefunction probes highly oscillatory potential energy curves and non-adiabatic couplings.
Surprisingly, the non-adiabatic transition rates for high energy dynamics (when the wavefunction start from the dissociation threshold), is qualitatively very well predicted by the LZ-model (see \cite{srikumar2025}).  This could be because high lying states and their momentum $k(R)$ can be reasonably approximated semi-classically, and their rapidly changing wavefunctions allow a WKB style representation, and probes the crossing with a well defined velocity. Even so, the standard Landau-Zener theory does not give us exact time evolving wavefunctions, as we aim to study here. It is of course possible that an exact multi-channel calculation which involves the degenerate manifolds,  reproduces the semiclassical result. But for now we continue presenting the results of our two channel study. 
\\

Figures \ref{fig:diabatic}(a) and \ref{fig:diabatic}(b) show the diabatic PECs  and the adiabatic PECs (solid black curves) for $n=55$ and $n=57$ respectively. Primarily, we see that the highly oscillatory (diagonal) diabatic PECs start deviating from their adiabatic counterparts, far away from the avoided crossing ($>1000\mathrm{a}_0$). This is accompanied by the off-diagonal potentials acquiring large non-zero values away from the avoided crossing. Moreover, we see that the global structure of the diabatic curves look reasonably similar between $n=55$ and $n=57$, despite a large difference in the magnitude of derivative couplings. However, the magnified plots (insets of Fig.~\ref{fig:diabatic}) do show that adiabatic PECs exhibit a narrow near-exact crossing for $n=55$, and a wider avoided crossing for $n=57$. This validates the large difference in the magnitude of $P_{12}$ for the two $n$-values. To summarize, the non-adiabatic couplings, and the diabatic potentials of the two-channel system deviate considerably from traditional molecular physics, or Landau-Zener models. A more detailed analysis of the vibronic couplings between trilobite and butterfly potentials can be found elsewhere (see Ref.~\cite{Hummel_2022}). Note that the exact structure of the diabatic potentials varies considerably depending on the number of channels used in diabatization. In reality, the highly polar electronic states are non-adiabatically coupled to other quantum defect states, as well as the quasi-degenerate manifold. To explain higher dynamics spanning the entire potential depth, one would need to incorporate higher quantum defect states and the unperturbed high-$\ell$ manifold into a multichannel model. Fortunately, the internal diffraction dynamics we probe requires relatively lower kinetic energies \cite{srikumar2024}. Since we see that the non-adiabatic couplings are still maximal at the avoided crossing, a two-channel model that considers the avoided crossing should give us a reliable semi-quantitative model for the problem at hand.

\begin{figure}
    \centering
    \includegraphics[width=\linewidth]{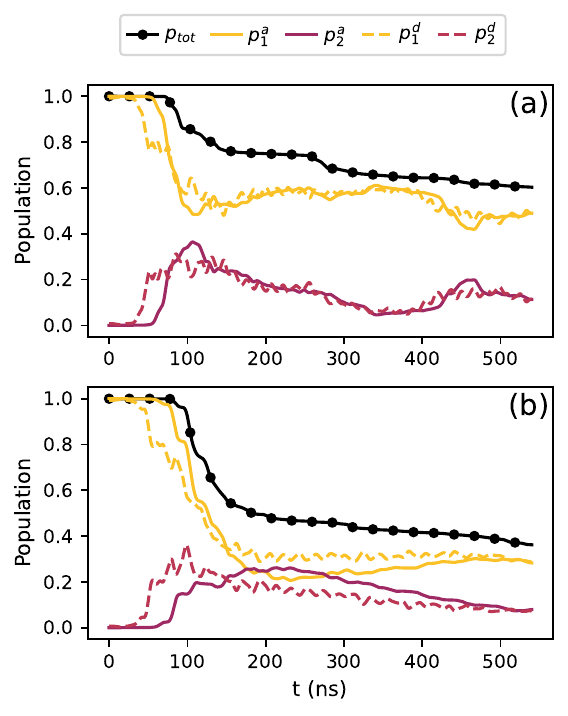}
    \caption{Time evolution of the diabatic  populations $p_{d}$ (dashed curves), and the adiabatic populations $p_{a}$ (solid curves) for channel 1 and 2, as well as the the total population $p_{tot}$ (dotted curve). (a) Depicts the population dynamics for $n=55$, and (b) depicts the population dynamics for $n=57$. }
    \label{fig:population}
\end{figure}

\subsection{Two-channel wavepacket dynamics} \label{Sec3B}

We begin our simulation of the wave-packet dynamics, by initializing the nuclear wavefunction as a Gaussian state in the $V_1$ adiabatic PEC, corresponding to the $\psi_1$ electronic state.
The initial position and width of the Gaussian are taken to be so that the ensuing dynamics display internal diffraction effects. This is because the oscillatory structure of the trilobite potential acts as a diffraction grating in momentum space. The anharmonic envelope of the pure trilobite PEC allows the wavepacket to gain momentum as it traverses down the potential, consequently scattering off the oscillatory structure of the PEC. At appropriate width and momenta, the consequent coherent back-scattering causes the wavepacket to appear as a propagating diffraction pattern. A comprehensive analysis of this phenomena is presented elsewhere (see Ref.~\cite{srikumar2024}),including the appropriate initial parameters needed to observe this effect. Here we build upon the previous work, and take the required parameters needed to observe diffraction, as given. Our aim then, is to see how the vibronic coupling between trilobite and potential affects the previously simulated diffraction dynamics. We try to analyze the $n$-dependence of the vibronic corrections, by contrasting the $n=55$ case with $n=57$, featuring the most and the least non-adiabatic coupling strength at the avoided crossing, respectively.

The relevant potential energy structure of the $n=55$ and $n=57$ molecule are shown in figure \ref{fig:probability_density}(a) and (b). We choose to initialize the Gaussian wavepacket $\chi_1^a(t=0)$ (Eq.~\ref{Eq:vibronic_chi}), with $\sigma= 100 \mathrm{a}_0$, and  $R_{0}= 4000 \mathrm{a}_0$. The width $\sigma$ is chosen while taking into consideration the available initial states (vibrational ground states of low-$\ell$ PECs \cite{srikumar2024}) for the pump-probe experiment. At the given $R_0$ the diabatic and adiabatic PECs coincide and show a predominant trilobite character. Moreover, the initial configuration presented was also used in the previous single channel study \cite{srikumar2024}, giving the results comparative value.  Figure \ref{fig:probability_density} (c) and (d) show the resulting time evolving probability density functions for $n=55$ and $n=57$, in position representation. As expected, simulations with both the $n$-values exhibit a propagating diffraction pattern for $t<100$ nanoseconds, until they probe the avoided crossing. We see that $n=55$ continues to exhibit propagating wavepacket dynamics mimicking the pure trilobite case, although there is a noticeable decrease in the contrast of the pattern compared to the single channel case. This can be explained by the consistent non-adiabatic transition of the nuclear wavepacket between $V_1$ and $V_2$ PECs due to the near exact crossing, so that the nuclear wavepacket effectively experience a pure trilobite potential, despite the actual wavepacket dynamics occurring on coupled channels. On the other hand, the $n=57$ case, with a wider avoided crossing, show a much clearer diminishing of the diffraction pattern after the first crossing. As the wavepacket keeps oscillating it probes the crossing multiple times, thereby destroying the diffraction pattern. When analyzing the dynamics in the momentum space representation (see figure \ref{fig:probability_density} (e) and (f)),  we see that the diffraction pattern is more coherent for the the $n$-value with the more narrow crossing. In addition, when the wavepacket reaches the avoided crossing for the first time, we see that a considerable fraction of the wavepacket gain momentum and escape the propagating diffraction pattern. This is because the nuclei which follow the $P$-wave decay pathway, experience a steep drop in the potential, resulting in the wavefunction gaining more momentum before it is eventually absorbed by the CAP. This phenomena is more dramatically observed in the case of $n=57$ due to the wider avoided crossing facilitating more adiabatic decay.  \\

\begin{figure*}
    \centering
    \includegraphics[width=\linewidth]{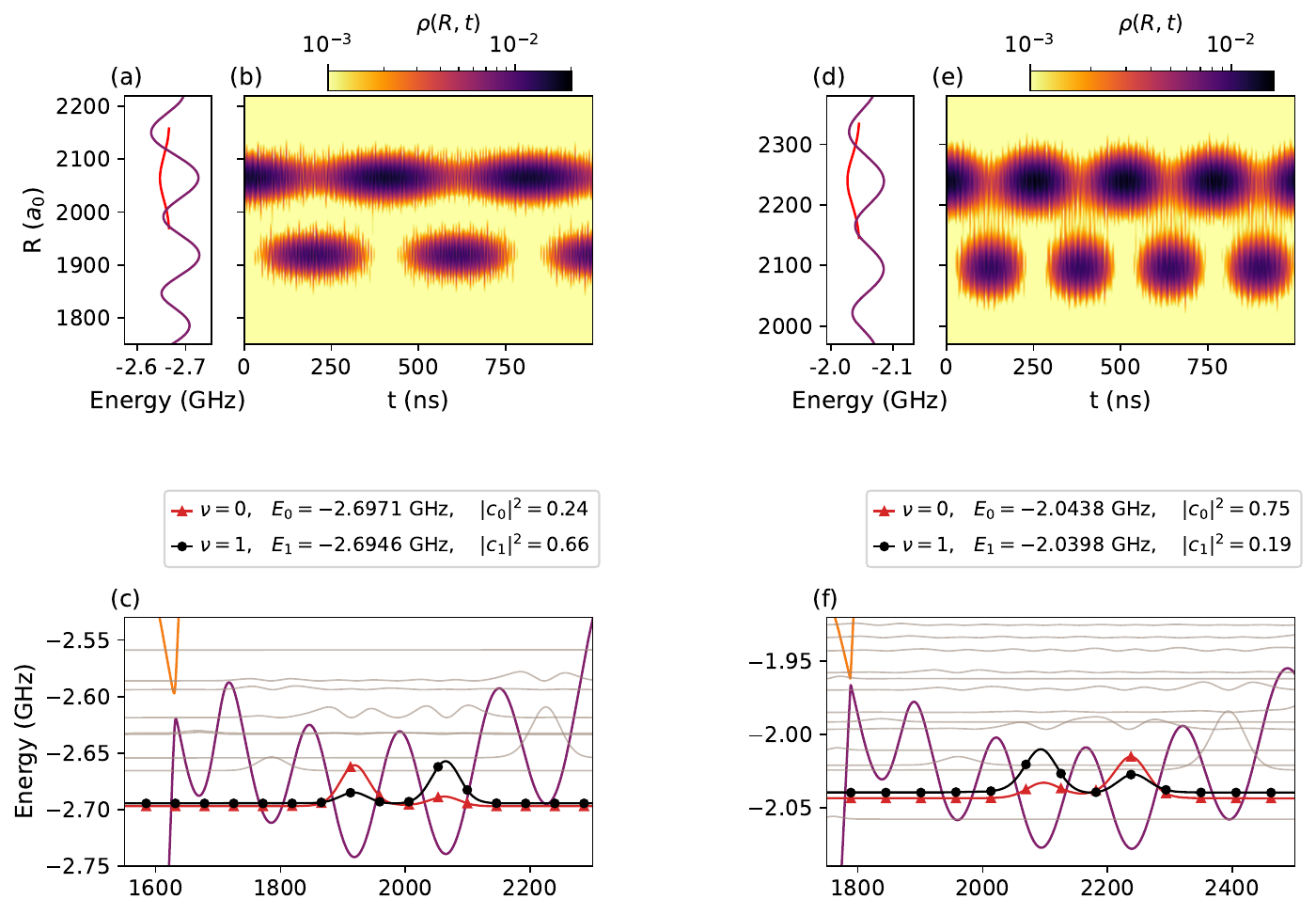}
    \caption{Two-well interference effects in the low energy dynamics of (a-c) $n=42$, and (d-f) $n=46$ ULRM. (a) and (d) Show the adiabatic potential energy curve $V_1$ for the ULRM, along with probability density of the initial Gaussian wavepacket with $\sigma= 40\mathrm{a}_0$, centered around $R_0=2064 \mathrm{a}_0$ and $R_0=2240 \mathrm{a}_0$ respectively. (b) and (e) Show the time evolving total probability density $\rho(R,t)$. (c) and (f) Display the vibronic molecular resonances.  The two bound eigenstates responsible for the observed dynamics are highlighted and labeled, along with their energies $E_{\nu}$, and their contribution to the wavepacket $|c_{\nu}|^2=|\langle \chi(R,t)| \chi_{\nu}\rangle|^2$. }
    \label{fig:tunnel2}
\end{figure*}

We now focus on the dynamics of the population calculated using aforementioned formulae (Eq.~\ref{eq:pop_dia} - Eq.~\ref{eq:pop_tot}). Figure \ref{fig:population} (a) and (b) shows the total population (marked $\bullet$), as well as the channel populations in adiabatic (solid lines) and diabatic (dashed lines) representations. For both $n$-values, the total population is constant (unit value) for the first 100 nanoseconds, until the wavefunction probes the avoided crossing. The population starts decreasing as expected, when the wavepacket is absorbed by the CAP after the avoided crossing. We also see that the population in the adiabatic channels start mixing slightly before the total population decrease. This corresponds to the short time dynamics that characterize the transfer of population from the trilobite to the butterfly state near the avoided crossing. The diabatic channel populations however, start mixing much before the wavepackets reach the avoided crossing, in accordance with our analysis of extended off-diagonal non-adiabatic couplings in the previous section. This hints at the non Landau-Zener nature of the mathematical model used. Comparing figure \ref{fig:population} (a) to (b), we see that the decrease in total population is indeed more prominent for $n=57$, due to the wider avoided crossing.

\begin{figure*}
    \centering
    \includegraphics[width=\linewidth]{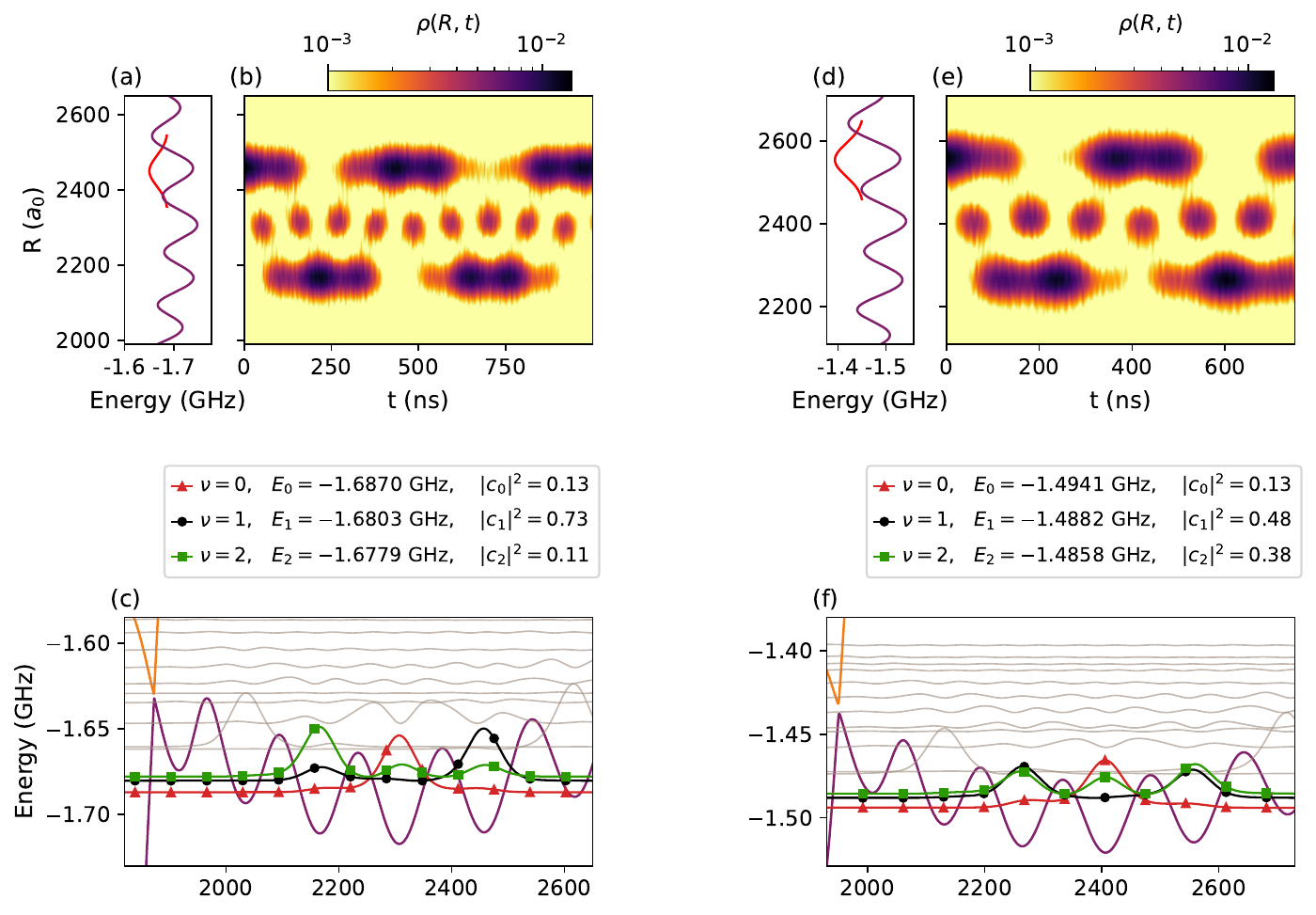}
    \caption{Three-well interference effects in the low energy dynamics of (a-c) $n=49$, and (d-f) $n=51$ ULRM. (a) and (d) show the adiabatic potential energy curve $V_1$ for the ULRM, along with probability density of the initial Gaussian wavepacket with $\sigma= 40\mathrm{a}_0$, centered around $R_0=2450 \mathrm{a}_0$ and $R_0=2555 \mathrm{a}_0$ respectively. (b) and (e) show the time evolving total probability density $\rho(R,t)$. (c) and (f) Display the vibronic molecular resonances.  The two bound eigenstates responsible for the observed dynamics are highlighted and labeled, along with their energies $E_{\nu}$, and their contribution to the wavepacket $|c_{\nu}|^2=|\langle \chi(R,t)| \chi_{\nu}\rangle|^2$. }
    \label{fig:tunnel3}
\end{figure*}

\subsection{Multi-well tunneling dynamics} \label{Sec3C}

Finally, we move on from higher energy vibronic dynamics, and shift our focus to low energy dynamics between different adjacent wells of PEC. We know that for low $n$-values ($\leq30$) vibrational ground states are generally localized in each single well of the oscillatory potential (see appendix of ref.~\cite{Exner2024}). Here, narrow Gaussian wavepackets (comparable to accessible pump states) that are initialized in any particular well, primarily perform intra-well oscillations. As we increase the $n$-value, the oscillatory structure becomes more and more shallow, decreasing the potential barrier between adjacent wells. Consequently, vibrational ground states near the trilobite minima start to become delocalized across multiple adjacent wells, and corresponding wavepackets exhibit multi-well tunneling effects.
\\

Starting from $n \geq 42$, the vibrational ground states can be delocalized across two wells, allowing us to engineer two-well interference effects.  Figure \ref{fig:tunnel2} (a) shows the adiabatic potential $V_1$ of the $n=42$, near the trilobite energetic minima.
We probe this region by initializing the vibrational wavepacket (red curve) as a Gaussian centered around the midpoint of the well   ($R_0=2064 \mathrm{a}_0$) , with $\sigma=40 \mathrm{a}_0$ for $n=42$. 
The ensuing dynamics is shown in figure \ref{fig:tunnel2} (b), where we see a periodic interference pattern that spans two different wells.  The pattern can be explained by decomposing the wavepacket in the vibronic eigenbasis, only considering eigenstates with a large contribution ($|c_{\nu}|^2=|\langle \chi(R,t)| \chi_{\nu}\rangle|^2$).
The vibronic spectra are plotted in figure \ref{fig:tunnel2} (c), where we see that the two highlighted wavefunctions (the ground state and the first excited state of the trilobite PEC), delocalized across the two potential wells, possess the highest overlap with the initial states. The other states only marginally contribute to the dynamics (grey lines), due to negligible overlap with the initial wavefunction.
One can then successfully explain the pattern as a two-well hopping dynamics, with a fixed period $\tau =400$ ns that is the inverse of the vibrational splitting $E_2 - E_1 = 2.5$ MHz. Similarly, figures
\ref{fig:tunnel2} (d-f) display and explain the two-well hopping for the $n=46$ ULRM. Just like the previous case, the wavepacket is a time evolving superposition of the two lowest energy vibration states, localized across two wells. Their larger vibrational splitting ($\sim 4$ MHz) creates a shorter hopping period of 250 ns.\\

For higher $n$-values in the range 49 to 52, the vibrational states can be localized across three different wells. Figure \ref{fig:tunnel3} (a) shows the adiabatic potential $V_1$ of the $n=49$ trilobite potential near its energetic minima, as well as the initial Gaussian wavepacket ($R_0 = 2450\mathrm{a}_0$), with $\sigma=40 \mathrm{a}_0$) of choice.  The ensuing dynamics is portrayed in \ref{fig:tunnel3} (b), where we see that pattern exhibits two different periods, with most of the population tunneling between next-nearest neighbouring wells. But some of the population periodically localizes in the intermediate well with a pulsing period almost twice as large as the tunneling period. The pattern can be mainly explained using the phase relations between the responsible eigenstates (solid lines in \ref{fig:tunnel3} (c) marked as $\nu=0,1,2$), that have a high overlap with the initial Gaussian.  While the highlighted bound vibronic states are centered around separate adiabatic potential wells, the wavefunction is clearly localized across other neighboring wells. This partial delocalized structure, along with very low level splitting leads to very interesting multi-well tunneling dynamics. The three eigenstates with the largest overlap with the initial wavepacket have energies $E_0 = -1.687$ GHz, $E_1 = -1.680$ GHz and $E_2 = -1.678$ GHz. The slow tunneling between the two outer wells is mainly governed by the interference between the first and second excited states. Their small splitting of $E_2-E_1 = 0.002$ GHz corresponds to an interference frequency of about $2$ MHz, giving rise to a tunneling timescale of roughly $500$ ns. In contrast, the faster pulsation of probability density in the central well arises mainly from the interference between the ground and second excited states. The larger splitting of $E_2-E_0 = 0.009$ GHz corresponds to a frequency of $9$ MHz, which explains the shorter modulation period of about $110$ ns visible in figure \ref{fig:tunnel3} (b). Figures \ref{fig:tunnel3} (d-f) present similar multi-well tunneling effects for the $n$=51 ULRM, whose origin can be traced back to three similarly  delocalized vibronic states. 
For even higher $n$-values we can also see four-well and five-well dynamical effects. However, since the wavefunctions are no longer properly localized around any single well, the emerging patterns become less clear. For increasing $n$, the oscillatory structure eventually becomes too shallow to hold localized vibrational states, so that multi-well tunneling effects are no longer visible.

\section{Conclusion and outlook} \label{Sec4}

Our work presents the first step in the study of multichannel vibronic dynamics in ULRM. We started by constructing a two-channel diabatic model from the coupled trilobite-butterfly system, and then proceed to numerically simulate the propagation of nuclear wavepackets using a Crank-Nicolson scheme. The wavefunction was initialized as a Gaussian, with appropriate mean and variance needed to facilitate previously studied diffraction effects. Once the wavepacket escapes into the butterfly channel and reaches small $R$-values, CAPs were used to simulate an absorbing wall (emulating internal molecular decay). 
We observed the two-channel analog of the diffraction effect with $n$-dependent non-adiabatic corrections. For $n$-values with extremely narrow avoided crossings, the propagating diffraction pattern is non-adiabatically stabilized against decay. For $n$-values with wide avoided crossings, larger fractions of the nuclear wavepacket is allowed to escape adiabatically. However, the high level density and delocalized non-adiabatic coupling suggest that high energy oscillations might need more than two-channels for accurate simulation. In the case of very low-energy dynamics, accurately simulated using a few single channel vibrational states, we observe interesting multi-well interference patterns due to the partially delocalized vibronic spetrum. \\

In principle, the results obtained here should be experimentally observable using state-of-the-art pump-probe schemes, as discussed in our previous work \cite{srikumar2024}. The preparation of the initial states, as well as certification of the dynamical effects can be performed using bound states in low-$\ell$ ULRM. Moreover, the two-channel model provided can also be used to simulate pre-dissociation dynamics and $\ell$-changing collisions \cite{Pfau2016,Hummel_2020} with low-$\ell$ quantum defect states. With the recent advancements in optical tweezer technology that now allows the individual assembly of Rydberg molecules \cite{Guttridge2024}, one can also attain the ability to prepare initial states as localized ground states with tunable separation. In such cases, modern three-photon excitation schemes allow us to
directly access the trilobite state  \cite{Exner2024,Althoen2023} through the $nf$-state admixture, with  MHz-scale laser linewidths. This allows us to project wavepackets directly onto the trilobite PEC, without worrying about off-resonant excitations to other potentials. 

From a theoretical perspective, a more ambitious extension of this work is to develop a multichannel model, considering the effect of the quasi-degenerate manifold, as well as couplings to nearby $n$-values. This will allow us to simulate vibronic dynamics in high-$\ell$ and low-$\ell$ molecules with quantitative rigor. Future theoretical investigations can also consider introducing an external electric or magnetic field, breaking the spherical symmetry of the systems and facilitating conical intersections. This could possibly result in interesting dynamical effects induced by geometric phases.

\section{Acknowledgements} \label{Sec5}

R.S. is grateful to Matthew T. Eiles and Hans Dieter Meyer for helpful discussions.

\begin{figure*}
    \centering
    \includegraphics[width=\linewidth]{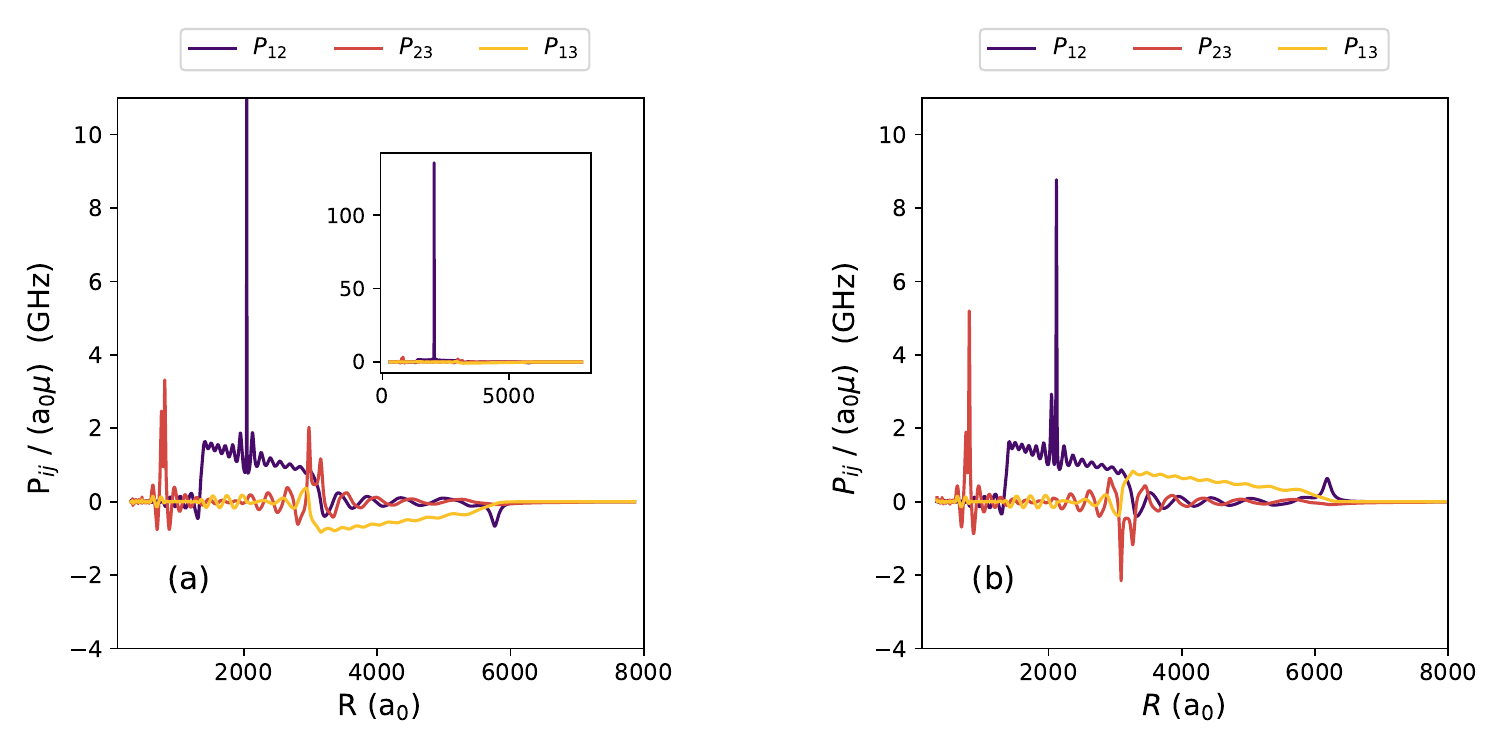}
    \caption{ The first derivative couplings $P_{ij}/(\mu \mathrm{a}_0)$ for (a) $n$=55 and (b) $n$=57. The insets in (a) are meant to show the near-singular behavior of the $n$=55 $P_{12}$, which is not necessary for $n$=57.}
    \label{fig:threechan}
\end{figure*}

\appendix
\section{Complex absorbing potentials} \label{A1}

Following the approach outlined in \cite{mctdh_intro, Riss1993}, the CAP is implemented as a negative imaginary potential of the form
\begin{align}
    &V_\mathrm{CAP}(R) = - i \, \eta \, W(R) \, , \\
    &W(R) = [R - R_c]^n \, \Theta(R_c - R) \, ,
\end{align}
where $\eta$ denotes the absorption strength, $R_c$ the onset position below which the CAP is active, $n$ the polynomial order, and $\Theta$ the Heaviside step function. Typical choices of $n=2$–$4$ yield a sufficiently smooth onset while keeping reflections minimal. Within this region, the total Hamiltonian becomes non-Hermitian,
\begin{equation}
    H' = H - i \eta W(R),
\end{equation}
which causes a decay of the total norm according to \cite{mctdh_intro}
\begin{equation}
    \frac{d}{dt} \lVert \Psi \rVert^2 = - 2 \eta \langle \Psi | W(R) | \Psi \rangle.
\end{equation}
Hence, the decrease of $\lVert \Psi \rVert^2$ directly reflects the loss of flux into the absorbing region.

For ultralong-range Rydberg molecules, optimal CAP parameters depend on the typical kinetic energy of the nuclear wavepacket. A sufficiently weak absorber ($\eta \sim 10^{-11}$ atomic units, or $\sim 10^{-5}$ GHz)  ensures that the physical dynamics in the interaction region remain unaffected while reflections are minimized. 
For each $n$-value, $R_c$ is chosen so that the effective CAP length ($\sim1000 \mathrm{a}_0$) far exceeds the local de Broglie wavelengths of the outgoing component to guarantee smooth absorption. In our two-channel model, the CAP is initially defined in the adiabatic representation, where it acts only within the butterfly potential channel. Upon transformation into the diabatic basis, the CAP is distributed over both diabatic components according to
\begin{equation}
    \boldsymbol{V}_\mathrm{CAP}^{(d)}(R) = U^\dagger(R) 
    \begin{pmatrix}
        V_\mathrm{CAP}(R) & 0 \\
        0 & 0
    \end{pmatrix}
    U(R),
\end{equation}
ensuring consistent damping of the diabatic wavepacket amplitudes.

\section{Three-channel derivative couplings} \label{A2}

Figure~\ref{fig:threechan} show the off-diagonal derivative couplings between the three adiabatic states $\psi_1,\psi_2$ and $\psi_3$, with the corresponding adiabatic PECs shown in Fig.~\ref{fig:potential_energy}.
The adiabatic states considered are formed by the mixing of the $f$-state, trilobite state, and the butterfly state. 
All the derivative coupling elements show highly oscillatory behavior, as they are comprised of hybridized radial atomic wavefunctions and their derivatives. In the vicinity of the trilobite-butterfly avoided crossing, $P_{12}$ dominates the other coupling terms; especially for the $n$=55 case, where its magnitude tis several orders of magnitude higher than the other coupling elements. We also see two peaks in $P_{23}$ that correspond to the avoided crossing between trilobite and $f$-state ($R\sim 800$ a$_0$) as well as the avoided crossing between butterfly and $f$-state ($R\sim 3000$ a$_0$).
Since we initialize our wavepacket in channel 1, the coupling between $\psi_2$ and $\psi_3$ are not so important for low energy oscillations, given that the wavepacket does not reach the $f$-state. However, we note that the coupling between channel 1 and 3 (given by $P_{13}$), is not negligible. In fact, $P_{13}$ seems to dominate $P_{12}$ at larger $R$-values beyond the crossing between butterfly and $f$-state ($R \geq 3000$ a$_0$). This can be explained by the change of $\psi_3$ from $f$-state to butterfly state beyond the avoided crossing, so that $P_{13}$ represents the coupling between trilobite and butterfly state at larger $R$-values. Moreover, all the off-diagonal terms show GHz-scale energy structure for an extended $R$-range away from the avoided crossing. This confirms that a quantitatively accurate treatment of the vibronic dynamics of ULRM, should definitely include three or more channels. But it is important to note that the non-adiabatic transition probability is also dependent on the particle momentum (or derivative of the wavefunction) and the energy splitting, not just $P_{ij}$. So the primary correction to the adiabatic approximation, and consequently the primary cause of $n$-dependent non-adiabatic stabilization, is the behavior of $P_{12}$ at the avoided crossing. \\

\bibliographystyle{apsrev4-2}
\bibliography{references}

@article{srikumar2024,
  title={Internal diffraction dynamics of trilobite molecules},
  author={Srikumar, Rohan and Rittenhouse, Seth T and Schmelcher, Peter},
  journal={Phys. Rev. A},
  volume={110},
  number={6},
  pages={062808},
  year={2024},
  publisher={APS},
  url={https://doi.org/10.1103/PhysRevA.110.062808}
}

@article{Greene2000,
  title = {Creation of Polar and Nonpolar Ultra-Long-Range Rydberg Molecules},
  author = {Greene, Chris H. and Dickinson, A. S. and Sadeghpour, H. R.},
  journal = {Phys. Rev. Lett.},
  volume = {85},
  issue = {12},
  pages = {2458--2461},
  numpages = {0},
  year = {2000},
  month = {Sep},
  publisher = {American Physical Society},
  doi = {10.1103/PhysRevLett.85.2458},
  url = {https://link.aps.org/doi/10.1103/PhysRevLett.85.2458}
}

@article{Bendkowsky2010,
  title = {Rydberg Trimers and Excited Dimers Bound by Internal Quantum Reflection},
  author = {Bendkowsky, V. and Butscher, B. and Nipper, J. and Balewski, J. B. and Shaffer, J. P. and L\"ow, R. and Pfau, T. and Li, W. and Stanojevic, J. and Pohl, T. and Rost, J. M.},
  journal = {Phys. Rev. Lett.},
  volume = {105},
  issue = {16},
  pages = {163201},
  numpages = {4},
  year = {2010},
  month = {Oct},
  publisher = {American Physical Society},
  doi = {10.1103/PhysRevLett.105.163201},
  url = {https://link.aps.org/doi/10.1103/PhysRevLett.105.163201}
}

@article{Hummel2020,
author = {Christian Fey and Frederic Hummel and Peter Schmelcher},
title = {Ultralong-range {R}ydberg molecules},
journal = {Mol. Phys.},
volume = {118},
number = {2},
pages = {e1679401},
year  = {2020},
publisher = {Taylor \& Francis},
doi = {10.1080/00268976.2019.1679401},

URL = { https://doi.org/10.1080/00268976.2019.1679401},


}

@article{Hummel_2022,
  title = {Vibronic interactions in trilobite and butterfly Rydberg molecules},
  author = {Hummel, Frederic and Schmelcher, Peter and Eiles, Matthew T.},
  journal = {Phys. Rev. Res.},
  volume = {5},
  issue = {1},
  pages = {013114},
  numpages = {11},
  year = {2023},
  month = {Feb},
  publisher = {American Physical Society},
  doi = {10.1103/PhysRevResearch.5.013114},
  url = {https://link.aps.org/doi/10.1103/PhysRevResearch.5.013114}
}

@article{Raithel2022,
  title = {Nonadiabatic decay of {R}ydberg-atom--ion molecules},
  author = {Duspayev, A. and Raithel, G.},
  journal = {Phys. Rev. A},
  volume = {105},
  issue = {1},
  pages = {012810},
  numpages = {8},
  year = {2022},
  month = {Jan},
  publisher = {American Physical Society},
  doi = {10.1103/PhysRevA.105.012810},
  url = {https://link.aps.org/doi/10.1103/PhysRevA.105.012810}
}

@article{Pfau2016,
  title = {Ultracold Chemical Reactions of a Single {R}ydberg Atom in a Dense Gas},
  author = {Schlagm\"uller, Michael and Liebisch, Tara Cubel and Engel, Felix and Kleinbach, Kathrin S. and B\"ottcher, Fabian and Hermann, Udo and Westphal, Karl M. and Gaj, Anita and L\"ow, Robert and Hofferberth, Sebastian and Pfau, Tilman and P\'erez-R\'{\i}os, Jes\'us and Greene, Chris H.},
  journal = {Phys. Rev. X},
  volume = {6},
  issue = {3},
  pages = {031020},
  numpages = {14},
  year = {2016},
  month = {Aug},
  publisher = {American Physical Society},
  doi = {10.1103/PhysRevX.6.031020},
  url = {https://link.aps.org/doi/10.1103/PhysRevX.6.031020}
}

@Article{Zuber2022,
author={Zuber, Nicolas
and Anasuri, Viraatt S. V.
and Berngruber, Moritz
and Zou, Yi Quan
and Meinert, Florian
and L{\"o}w, Robert
and Pfau, Tilman},
title={Observation of a molecular bond between ions and {R}ydberg atoms},
journal={Nature},
year={2022},
month={May},
day={01},
volume={605},
number={7910},
pages={453-456},
issn={1476-4687},
doi={10.1038/s41586-022-04577-5},
url={https://doi.org/10.1038/s41586-022-04577-5}
}

@article{Samboy2011,
  title = {Formation and properties of Rydberg macrodimers},
  author = {Samboy, Nolan and Stanojevic, Jovica and C\^ot\'e, Robin},
  journal = {Phys. Rev. A},
  volume = {83},
  issue = {5},
  pages = {050501(R)},
  numpages = {4},
  year = {2011},
  month = {May},
  publisher = {American Physical Society},
  doi = {10.1103/PhysRevA.83.050501},
  url = {https://link.aps.org/doi/10.1103/PhysRevA.83.050501}
}

@article{hollerith_2019,
author = {Simon Hollerith  and Johannes Zeiher  and Jun Rui  and Antonio Rubio-Abadal  and Valentin Walther  and Thomas Pohl  and Dan M. Stamper-Kurn  and Immanuel Bloch  and Christian Gross },
title = {Quantum gas microscopy of {R}ydberg macrodimers},
journal = {Science},
volume = {364},
number = {6441},
pages = {664-667},
year = {2019},
doi = {10.1126/science.aaw4150},
URL = {https://www.science.org/doi/abs/10.1126/science.aaw4150},

}

@article{Hollerith_2021,
  title = {Microscopic electronic structure tomography of Rydberg macrodimers},
  author = {Hollerith, Simon and Rui, Jun and Rubio-Abadal, Antonio and Srakaew, Kritsana and Wei, David and Zeiher, Johannes and Gross, Christian and Bloch, Immanuel},
  journal = {Phys. Rev. Res.},
  volume = {3},
  issue = {1},
  pages = {013252},
  numpages = {11},
  year = {2021},
  month = {Mar},
  publisher = {American Physical Society},
  doi = {10.1103/PhysRevResearch.3.013252},
  url = {https://link.aps.org/doi/10.1103/PhysRevResearch.3.013252}
}

@article{Engel_2019,
  title = {Precision Spectroscopy of Negative-Ion Resonances in Ultralong-Range {R}ydberg Molecules},
  author = {Engel, F. and Dieterle, T. and Hummel, F. and Fey, C. and Schmelcher, P. and L\"ow, R. and Pfau, T. and Meinert, F.},
  journal = {Phys. Rev. Lett.},
  volume = {123},
  issue = {7},
  pages = {073003},
  numpages = {6},
  year = {2019},
  month = {Aug},
  publisher = {American Physical Society},
  doi = {10.1103/PhysRevLett.123.073003},
  url = {https://link.aps.org/doi/10.1103/PhysRevLett.123.073003}
}

@article{Engel2024,
  title = {In situ observation of chemistry in Rydberg molecules within a Bose-Einstein-condensate},
  author = {Engel, Felix and Tiwari, Shiva Kant and Pfau, Tilman and W\"uster, Sebastian and Meinert, Florian},
  journal = {Phys. Rev. Res.},
  volume = {6},
  issue = {3},
  pages = {033150},
  numpages = {11},
  year = {2024},
  month = {Aug},
  publisher = {American Physical Society},
  doi = {10.1103/PhysRevResearch.6.033150},
  url = {https://link.aps.org/doi/10.1103/PhysRevResearch.6.033150}
}

@article{Anderson2014,
  title = {Angular-momentum couplings in long-range ${\mathrm{Rb}}_{2}$ Rydberg molecules},
  author = {Anderson, D. A. and Miller, S. A. and Raithel, G.},
  journal = {Phys. Rev. A},
  volume = {90},
  issue = {6},
  pages = {062518},
  numpages = {8},
  year = {2014},
  month = {Dec},
  publisher = {American Physical Society},
  doi = {10.1103/PhysRevA.90.062518},
  url = {https://link.aps.org/doi/10.1103/PhysRevA.90.062518}
}

@article{Born_1927,
author = {Born, M. and Oppenheimer, R.},
title = {Zur Quantentheorie der Molekeln},
journal = {Ann. Phys. (Leipzig)},
volume = {389},
number = {20},
pages = {457-484},
doi = {https://doi.org/10.1002/andp.19273892002},
url = {https://onlinelibrary.wiley.com/doi/abs/10.1002/andp.19273892002},

year = {1927}
}

@Book{Huang_1954,
  booktitle={Dynamical Theory of Crystal Lattices},
  author={Max Born and K. Huang},
  year={1954},
  publisher = {Oxford University Press}

}

@article{Cherkani_2001,
doi = {10.1088/0953-4075/34/1/304},
url = {https://doi.org/10.1088/0953-4075/34/1/304},
year = {2001},
month = {jan},
publisher = {},
volume = {34},
number = {1},
pages = {49},
author = {M H Cherkani and F Brouillard and M Chibisov},
title = {Summation rules for Coulomb wavefunction
products},
journal = {J. Phys. B: At. Mol. Opt. Phys.},
}

@article{Ermolaev2000,
  title = {New Summation Rules for Coulomb Wave Functions},
  author = {Chibisov, M. and Ermolaev, A. M. and Brouillard, F. and Cherkani, M. H.},
  journal = {Phys. Rev. Lett.},
  volume = {84},
  issue = {3},
  pages = {451--454},
  numpages = {0},
  year = {2000},
  month = {Jan},
  publisher = {American Physical Society},
  doi = {10.1103/PhysRevLett.84.451},
  url = {https://link.aps.org/doi/10.1103/PhysRevLett.84.451}
}

@Book{koppel_1984,
author = {Köppel, H. and Domcke, W. and Cederbaum, L. S.},
publisher = {John Wiley \& Sons, Ltd},
isbn = {9780470142813},
title = {Multimode Molecular Dynamics Beyond the Born-Oppenheimer Approximation},
booktitle = {Advances in Chemical Physics},
pages = {59-246},
doi = {https://doi.org/10.1002/9780470142813.ch2},
url = {https://onlinelibrary.wiley.com/doi/abs/10.1002/9780470142813.ch2},
year = {1984}
}

@book{Domcke2004,
author = {Domcke, Wolfgang and Yarkony, David R and Köppel, Horst},
title = {Conical Intersections},
publisher = {WORLD SCIENTIFIC},
year = {2004},
doi = {10.1142/5406},
address = {},
edition   = {},
URL = {https://www.worldscientific.com/doi/abs/10.1142/5406}
}

@book{Baer_2006,
  title={Beyond Born-Oppenheimer: Electronic Non-adiabatic Coupling Terms and Conical Intersections},
  isbn = {9780471778912},
  publisher = {John Wiley \& Sons, Ltd},
  author={Michael Baer},
  year={2006},
  doi = {10.1002/0471780081},
  url = {https://onlinelibrary.wiley.com/doi/book/10.1002/0471780081}
}

@article{durst2024,
  title = {Nonadiabatic couplings as a stabilization mechanism in long-range Rydberg molecules},
  author = {Durst, Aileen A. T. and Simi\ifmmode \acute{c}\else \'{c}\fi{}, Milena and Abraham, Neethu and Eiles, Matthew T.},
  journal = {Phys. Rev. A},
  volume = {111},
  issue = {3},
  pages = {032811},
  numpages = {10},
  year = {2025},
  month = {Mar},
  publisher = {American Physical Society},
  doi = {10.1103/PhysRevA.111.032811},
  url = {https://link.aps.org/doi/10.1103/PhysRevA.111.032811}
}

@Article{Fermi1934,
author={Fermi, Enrico},
title={Sopra lo Spostamento per Pressione delle Righe Elevate delle Serie Spettrali},
journal={Nuovo Cim.},
url={https://link.springer.com/article/10.1007/BF02959829},
year={1934},
month={Mar},
day={01},
volume={11},
number={3},
pages={157-166},
issn={1827-6121}}

@article{Omont_1977,
	author = {{Omont, A.}},
	title = {On the theory of collisions of atoms in {R}ydberg states with neutral particles},
	DOI= "10.1051/jphys:0197700380110134300",
	url= "https://doi.org/10.1051/jphys:0197700380110134300",
	journal = {J. Phys. France},
	year = 1977,
	volume = 38,
	number = 11,
	pages = "1343-1359",
}

@article{Rosario2024,
  title = {Ultralong-range Cs-RbCs Rydberg molecules: Nonadiabaticity of dipole moments},
  author = {Mellado-Alcedo, David and Guttridge, Alexander and Cornish, Simon L. and Sadeghpour, H. R. and Gonz\'alez-F\'erez, Rosario},
  journal = {Phys. Rev. A},
  volume = {110},
  issue = {1},
  pages = {013314},
  numpages = {9},
  year = {2024},
  month = {Jul},
  publisher = {American Physical Society},
  doi = {10.1103/PhysRevA.110.013314},
  url = {https://link.aps.org/doi/10.1103/PhysRevA.110.013314}
}

@article{Rosario2021,
  title = {Ultralong-Range Rydberg Bimolecules},
  author = {Gonz\'alez-F\'erez, Rosario and Shertzer, Janine and Sadeghpour, H. R.},
  journal = {Phys. Rev. Lett.},
  volume = {126},
  issue = {4},
  pages = {043401},
  numpages = {5},
  year = {2021},
  month = {Jan},
  publisher = {American Physical Society},
  doi = {10.1103/PhysRevLett.126.043401},
  url = {https://link.aps.org/doi/10.1103/PhysRevLett.126.043401}
}

@article{Hamilton_2002,
doi = {10.1088/0953-4075/35/10/102},
url = {https://dx.doi.org/10.1088/0953-4075/35/10/102},
year = {2002},
month = {may},
publisher = {},
volume = {35},
number = {10},
pages = {L199},
author = {Edward L Hamilton and  Chris H Greene and  H R Sadeghpour},
title = {Shape-resonance-induced long-range molecular {R}ydberg
states},
journal = {J. Phys. B: At. Mol. Opt. Phys.}
}

@article{Riss1993,
doi = {10.1088/0953-4075/26/23/021},
url = {https://doi.org/10.1088/0953-4075/26/23/021},
year = {1993},
month = {dec},
publisher = {},
volume = {26},
number = {23},
pages = {4503},
author = {U V Riss and H -D Meyer},
title = {Calculation of resonance energies and widths using the complex absorbing potential method},
journal = {J. Phys. B: At. Mol. Opt. Phys}
}

@misc{mctdh_intro,
  author = {Meyer, Hans Dieter},
  title = {Introduction to {MCTDH}},
  howpublished = {Lecture Notes, Heidelberg University},
  url = {https://www.pci.uni-heidelberg.de/tc/usr/mctdh/lit/intro_MCTDH.pdf}
  }

@article{Feit1982,
title = {Solution of the Schrödinger equation by a spectral method},
journal = {J. Comput. Phys.},
volume = {47},
number = {3},
pages = {412-433},
year = {1982},
issn = {0021-9991},
doi = {https://doi.org/10.1016/0021-9991(82)90091-2},
url = {https://www.sciencedirect.com/science/article/pii/0021999182900912},
author = {M.D Feit and J.A Fleck and A Steiger},
}

@article{Borodin_1992,
doi = {10.1088/0953-4075/25/5/011},
url = {https://dx.doi.org/10.1088/0953-4075/25/5/011},
year = {1992},
month = {mar},
publisher = {},
volume = {25},
number = {5},
pages = {971},
author = {V M Borodin and A K Kazansky},
title = {The adiabatic mechanism of the collisional broadening of Rydberg states via a loosely bound resonance state of ambient gas atoms},
journal = {J. Phys. B: At. Mol. Opt. Phys.}
}

@article{Eiles_2017,
  title = {Hamiltonian for the inclusion of spin effects in long-range {R}ydberg molecules},
  author = {Eiles, Matthew T. and Greene, Chris H.},
  journal = {Phys. Rev. A},
  volume = {95},
  issue = {4},
  pages = {042515},
  numpages = {13},
  year = {2017},
  month = {Apr},
  publisher = {American Physical Society},
  doi = {10.1103/PhysRevA.95.042515},
  url = {https://link.aps.org/doi/10.1103/PhysRevA.95.042515}
}

@article{Hummel2021,
  title = {Synthetic Dimension-Induced Conical Intersections in Rydberg Molecules},
  author = {Hummel, Frederic and Eiles, Matthew T. and Schmelcher, Peter},
  journal = {Phys. Rev. Lett.},
  volume = {127},
  issue = {2},
  pages = {023003},
  numpages = {6},
  year = {2021},
  month = {Jul},
  publisher = {American Physical Society},
  doi = {10.1103/PhysRevLett.127.023003},
  url = {https://link.aps.org/doi/10.1103/PhysRevLett.127.023003}
}

@article{Kurz_2013,
  title = {Electrically dressed ultra-long-range polar {R}ydberg molecules},
  author = {Kurz, Markus and Schmelcher, Peter},
  journal = {Phys. Rev. A},
  volume = {88},
  issue = {2},
  pages = {022501},
  numpages = {8},
  year = {2013},
  month = {Aug},
  publisher = {American Physical Society},
  doi = {10.1103/PhysRevA.88.022501},
  url = {https://link.aps.org/doi/10.1103/PhysRevA.88.022501}
}

@article{Fey2019,
  title = {Effective Three-Body Interactions in $\mathrm{Cs}(6s)\text{\ensuremath{-}}\mathrm{Cs}(nd)$ Rydberg Trimers},
  author = {Fey, Christian and Yang, Jin and Rittenhouse, Seth T. and Munkes, Fabian and Baluktsian, Margarita and Schmelcher, Peter and Sadeghpour, H. R. and Shaffer, James P.},
  journal = {Phys. Rev. Lett.},
  volume = {122},
  issue = {10},
  pages = {103001},
  numpages = {5},
  year = {2019},
  month = {Mar},
  publisher = {American Physical Society},
  doi = {10.1103/PhysRevLett.122.103001},
  url = {https://link.aps.org/doi/10.1103/PhysRevLett.122.103001}
}

@article{Fey2016,
  title = {Stretching and bending dynamics in triatomic ultralong-range Rydberg molecules},
  author = {Fey, Christian and Kurz, Markus and Schmelcher, Peter},
  journal = {Phys. Rev. A},
  volume = {94},
  issue = {1},
  pages = {012516},
  numpages = {14},
  year = {2016},
  month = {Jul},
  publisher = {American Physical Society},
  doi = {10.1103/PhysRevA.94.012516},
  url = {https://link.aps.org/doi/10.1103/PhysRevA.94.012516}
}

@article{Wang2015,
  title = {Rydberg Electrons in a Bose-Einstein Condensate},
  author = {Wang, Jia and Gacesa, Marko and C\^ot\'e, R.},
  journal = {Phys. Rev. Lett.},
  volume = {114},
  issue = {24},
  pages = {243003},
  numpages = {5},
  year = {2015},
  month = {Jun},
  publisher = {American Physical Society},
  doi = {10.1103/PhysRevLett.114.243003},
  url = {https://link.aps.org/doi/10.1103/PhysRevLett.114.243003}
}

@article{Eiles_2019,
doi = {10.1088/1361-6455/ab19ca},
url = {https://dx.doi.org/10.1088/1361-6455/ab19ca},
year = {2019},
month = {may},
publisher = {IOP Publishing},
volume = {52},
number = {11},
pages = {113001},
author = {Matthew T Eiles},
title = {Trilobites, butterflies, and other exotic specimens of long-range {R}ydberg molecules},
journal = {J. Phys. B: At. Mol. Opt. Phys.}
}

@article{Hummel_2020,
doi = {10.1088/1367-2630/ab90d7},
url = {https://dx.doi.org/10.1088/1367-2630/ab90d7},
year = {2020},
month = {jun},
publisher = {IOP Publishing},
volume = {22},
number = {6},
pages = {063060},
author = {Hummel, F and Schmelcher, P and Ott, H and Sadeghpour, H R},
title = {An ultracold heavy Rydberg system formed from ultra-long-range molecules bound in a stairwell potential},
journal = {New J. Phys.}
}

@article{Keiler_2021,
  title = {Electric-field-induced wave-packet dynamics and geometrical rearrangement of trilobite {R}ydberg molecules},
  author = {Hummel, Frederic and Keiler, Kevin and Schmelcher, Peter},
  journal = {Phys. Rev. A},
  volume = {103},
  issue = {2},
  pages = {022827},
  numpages = {9},
  year = {2021},
  month = {Feb},
  publisher = {American Physical Society},
  doi = {10.1103/PhysRevA.103.022827},
  url = {https://link.aps.org/doi/10.1103/PhysRevA.103.022827}
}

@article{Groenenboom_1990,
author = {Groenenboom,Gerrit C.  and Buck,Henk M. },
title = {Solving the discretized time‐independent Schrödinger equation with the Lanczos procedure},
journal = {J. Chem. Phys.},
volume = {92},
number = {7},
pages = {4374-4379},
year = {1990},
doi = {10.1063/1.458575},

URL = { 
        https://doi.org/10.1063/1.458575
    
}
}

@article{anasuri_2023,
  title = {Observation of Vibrational Dynamics of Orientated Rydberg-Atom-Ion Molecules},
  author = {Zou, Yi Quan and Berngruber, Moritz and Anasuri, Viraatt S. V. and Zuber, Nicolas and Meinert, Florian and L\"ow, Robert and Pfau, Tilman},
  journal = {Phys. Rev. Lett.},
  volume = {130},
  issue = {2},
  pages = {023002},
  numpages = {6},
  year = {2023},
  month = {Jan},
  publisher = {American Physical Society},
  doi = {10.1103/PhysRevLett.130.023002},
  url = {https://link.aps.org/doi/10.1103/PhysRevLett.130.023002}
}

@article{bosworth_2022,
  title = {Charged ultralong-range Rydberg trimers},
  author = {Bosworth, Daniel J. and Hummel, Frederic and Schmelcher, Peter},
  journal = {Phys. Rev. A},
  volume = {107},
  issue = {2},
  pages = {022807},
  numpages = {10},
  year = {2023},
  month = {Feb},
  publisher = {American Physical Society},
  doi = {10.1103/PhysRevA.107.022807},
  url = {https://link.aps.org/doi/10.1103/PhysRevA.107.022807}
}

@article{Stanojevic2008,
  title = {Long-range potentials and $(n\ensuremath{-}1)d+ns$ molecular resonances in an ultracold Rydberg gas},
  author = {Stanojevic, J. and C\^ot\'e, R. and Tong, D. and Eyler, E. E. and Gould, P. L.},
  journal = {Phys. Rev. A},
  volume = {78},
  issue = {5},
  pages = {052709},
  numpages = {11},
  year = {2008},
  month = {Nov},
  publisher = {American Physical Society},
  doi = {10.1103/PhysRevA.78.052709},
  url = {https://link.aps.org/doi/10.1103/PhysRevA.78.052709}
}

@Article{Stanojevic2006,
author={Stanojevic, J.
and C{\^o}t{\'e}, R.
and Tong, D.
and Farooqi, S. M.
and Eyler, E. E.
and Gould, P. L.},
title={Long-range Rydberg-Rydberg interactions and molecular resonances},
journal={Eur. Phys. J. D},
year={2006},
month={Oct},
day={01},
volume={40},
number={1},
pages={3-12},
issn={1434-6079},
doi={10.1140/epjd/e2006-00143-x},
url={https://doi.org/10.1140/epjd/e2006-00143-x}
}

@Article{Niederprum2016,
author={Niederpr{\"u}m, Thomas
and Thomas, Oliver
and Eichert, Tanita
and Lippe, Carsten
and P{\'e}rez-R{\'i}os, Jes{\'u}s
and Greene, Chris H.
and Ott, Herwig},
title={Observation of pendular butterfly Rydberg molecules},
journal={Nat. Commun.},
year={2016},
month={Oct},
day={05},
volume={7},
number={1},
pages={12820},
issn={2041-1723},
doi={10.1038/ncomms12820},
url={https://doi.org/10.1038/ncomms12820}
}

@article{Lesanovsky_2006,
doi = {10.1088/0953-4075/39/4/L03},
url = {https://dx.doi.org/10.1088/0953-4075/39/4/L03},
year = {2006},
month = {feb},
publisher = {},
volume = {39},
number = {4},
pages = {L69},
author = {Igor Lesanovsky and Peter Schmelcher and H R Sadeghpour},
title = {Ultra-long-range Rydberg molecules exposed to a magnetic field},
journal = {J. Phys. B: At. Mol. Opt. Phys.}
}

@article{Wilkinson2024,
  title = {Spectral Signatures of Vibronic Coupling in Trapped Cold Ionic Rydberg Systems},
  author = {Wilkinson, Joseph W. P. and Li, Weibin and Lesanovsky, Igor},
  journal = {Phys. Rev. Lett.},
  volume = {132},
  issue = {22},
  pages = {223401},
  numpages = {7},
  year = {2024},
  month = {May},
  publisher = {American Physical Society},
  doi = {10.1103/PhysRevLett.132.223401},
  url = {https://link.aps.org/doi/10.1103/PhysRevLett.132.223401}
}

@article{Gambetta2021,
  title = {Exploring the Many-Body Dynamics Near a Conical Intersection with Trapped Rydberg Ions},
  author = {Gambetta, Filippo M. and Zhang, Chi and Hennrich, Markus and Lesanovsky, Igor and Li, Weibin},
  journal = {Phys. Rev. Lett.},
  volume = {126},
  issue = {23},
  pages = {233404},
  numpages = {6},
  year = {2021},
  month = {Jun},
  publisher = {American Physical Society},
  doi = {10.1103/PhysRevLett.126.233404},
  url = {https://link.aps.org/doi/10.1103/PhysRevLett.126.233404}
}

@article{Hummel_2019,
  title = {Alignment of $s$-state {R}ydberg molecules in magnetic fields},
  author = {Hummel, Frederic and Fey, Christian and Schmelcher, Peter},
  journal = {Phys. Rev. A},
  volume = {99},
  issue = {2},
  pages = {023401},
  numpages = {6},
  year = {2019},
  month = {Feb},
  publisher = {American Physical Society},
  doi = {10.1103/PhysRevA.99.023401},
  url = {https://link.aps.org/doi/10.1103/PhysRevA.99.023401}
}

@article{Herman_1984,
author = {Herman,Michael F. },
title = {Nonadiabatic semiclassical scattering. I. Analysis of generalized surface hopping procedures},
journal = {J. Chem. Phys.},
volume = {81},
number = {2},
pages = {754-763},
year = {1984},
doi = {10.1063/1.447708},

URL = { 
        https://doi.org/10.1063/1.447708
    
}


}

@article{Barbatti_2011,
author = {Barbatti, Mario},
title = {Nonadiabatic dynamics with trajectory surface hopping method},
journal = {Wiley Interdiscip. Rev. Comput. Mol. Sci.},
volume = {1},
number = {4},
pages = {620-633},
doi = {https://doi.org/10.1002/wcms.64},
url = {https://wires.onlinelibrary.wiley.com/doi/abs/10.1002/wcms.64},
year = {2011}
}

@Article{Shaffer2018,
author={Shaffer, J. P.
and Rittenhouse, S. T.
and Sadeghpour, H. R.},
title={Ultracold {R}ydberg molecules},
journal={Nat. Commun.},
year={2018},
month={May},
day={17},
volume={9},
number={1},
pages={1965},
issn={2041-1723},
doi={10.1038/s41467-018-04135-6},
url={https://doi.org/10.1038/s41467-018-04135-6}
}

@article{Butscher_2011,
doi = {10.1088/0953-4075/44/18/184004},
url = {https://dx.doi.org/10.1088/0953-4075/44/18/184004},
year = {2011},
month = {sep},
publisher = {},
volume = {44},
number = {18},
pages = {184004},
author = {Björn Butscher and Vera Bendkowsky and Johannes Nipper and Jonathan B Balewski and Ludmila Kukota and Robert Löw and Tilman Pfau and Weibin Li and Thomas Pohl and Jan Michael Rost},
title = {Lifetimes of ultralong-range {R}ydberg molecules in vibrational ground and excited states},
journal = {J. Phys. B: At. Mol. Opt. Phys.}
}

@article{Booth_2015,
author = {D. Booth  and S. T. Rittenhouse  and J. Yang  and H. R. Sadeghpour  and J. P. Shaffer },
title = {Production of trilobite Rydberg molecule dimers with kilo-Debye permanent electric dipole moments},
journal = {Science},
volume = {348},
number = {6230},
pages = {99-102},
year = {2015},
doi = {10.1126/science.1260722},
URL = {https://www.science.org/doi/abs/10.1126/science.1260722},
}

@article{srikumar2025,
  title = {Vibrationally highly excited trilobite molecules stabilized by nonadiabatic coupling},
  author = {Srikumar, Rohan and Exner, Markus and Bl\"attner, Richard and Schmelcher, Peter and Eiles, Matthew T. and Ott, Herwig},
  journal = {Phys. Rev. Res.},
  volume = {7},
  issue = {4},
  pages = {043103},
  numpages = {8},
  year = {2025},
  month = {Oct},
  publisher = {American Physical Society},
  doi = {10.1103/49ky-d8mm},
  url = {https://link.aps.org/doi/10.1103/49ky-d8mm}
}

@article{Sadeghpour_2011,
author = {W. Li  and T. Pohl  and J. M. Rost  and Seth T. Rittenhouse  and H. R. Sadeghpour  and J. Nipper  and B. Butscher  and J. B. Balewski  and V. Bendkowsky  and R. Löw  and T. Pfau },
title = {A Homonuclear Molecule with a Permanent Electric Dipole Moment},
journal = {Science},
volume = {334},
number = {6059},
pages = {1110-1114},
year = {2011},
doi = {10.1126/science.1211255},
URL = {https://www.science.org/doi/abs/10.1126/science.1211255}}

@article{Srikumar2023,
  title = {Nonadiabatic interaction effects in the spectra of ultralong-range Rydberg molecules},
  author = {Srikumar, Rohan and Hummel, Frederic and Schmelcher, Peter},
  journal = {Phys. Rev. A},
  volume = {108},
  issue = {1},
  pages = {012809},
  numpages = {12},
  year = {2023},
  month = {Jul},
  publisher = {American Physical Society},
  doi = {10.1103/PhysRevA.108.012809},
  url = {https://link.aps.org/doi/10.1103/PhysRevA.108.012809}
}

@Article{Peng2011,
author ="Zhang, Peng and Dalgarno, Alexander and Côté, Robin and Bodo, Enrico",
title  ="Charge exchange in collisions of beryllium with its ion",
journal  ="Phys. Chem. Chem. Phys.",
year  ="2011",
volume  ="13",
issue  ="42",
pages  ="19026-19035",
publisher  ="The Royal Society of Chemistry",
doi  ="10.1039/C1CP21494B",
url  ="http://dx.doi.org/10.1039/C1CP21494B"
}

@article{Killian2023,
  title = {Measuring nonlocal three-body spatial correlations with Rydberg trimers in ultracold quantum gases},
  author = {Kanungo, S. K. and Lu, Y. and Dunning, F. B. and Yoshida, S. and Burgd\"orfer, J. and Killian, T. C.},
  journal = {Phys. Rev. A},
  volume = {107},
  issue = {3},
  pages = {033322},
  numpages = {6},
  year = {2023},
  month = {Mar},
  publisher = {American Physical Society},
  doi = {10.1103/PhysRevA.107.033322},
  url = {https://link.aps.org/doi/10.1103/PhysRevA.107.033322}
}

@Article{Althoen2023,
author={Alth{\"o}n, Max
and Exner, Markus
and Bl{\"a}ttner, Richard
and Ott, Herwig},
title={Exploring the vibrational series of pure trilobite Rydberg molecules},
journal={Nat. Commun.},
year={2023},
month={Dec},
day={07},
volume={14},
number={1},
pages={8108},
issn={2041-1723},
doi={10.1038/s41467-023-43818-7},
url={https://doi.org/10.1038/s41467-023-43818-7}
}

@article{Gallagher2003,
  title = {Millimeter-wave spectroscopy of cold Rb Rydberg atoms in a magneto-optical trap: Quantum defects of the ns, np, and nd series},
  author = {Li, Wenhui and Mourachko, I. and Noel, M. W. and Gallagher, T. F.},
  journal = {Phys. Rev. A},
  volume = {67},
  issue = {5},
  pages = {052502},
  numpages = {7},
  year = {2003},
  month = {May},
  publisher = {American Physical Society},
  doi = {10.1103/PhysRevA.67.052502},
  url = {https://link.aps.org/doi/10.1103/PhysRevA.67.052502}
}

@article{eiles2024,
  title = {Kato's theorem and ultralong-range Rydberg molecules},
  author = {Eiles, Matthew T. and Hummel, Frederic},
  journal = {Phys. Rev. A},
  volume = {109},
  issue = {2},
  pages = {022811},
  numpages = {6},
  year = {2024},
  month = {Feb},
  publisher = {American Physical Society},
  doi = {10.1103/PhysRevA.109.022811},
  url = {https://link.aps.org/doi/10.1103/PhysRevA.109.022811}
}

@article{eiles2023,
  title = {Green's-function treatment of Rydberg molecules with spins},
  author = {Greene, Chris H. and Eiles, Matthew T.},
  journal = {Phys. Rev. A},
  volume = {108},
  issue = {4},
  pages = {042805},
  numpages = {18},
  year = {2023},
  month = {Oct},
  publisher = {American Physical Society},
  doi = {10.1103/PhysRevA.108.042805},
  url = {https://link.aps.org/doi/10.1103/PhysRevA.108.042805}
}

@article{Exner2024,
  title = {High Precision Spectroscopy of Trilobite Rydberg Molecules},
  author = {Exner, Markus and Srikumar, Rohan and Bl\"attner, Richard and Eiles, Matthew T. and Schmelcher, Peter and Ott, Herwig},
  journal = {Phys. Rev. Lett.},
  volume = {134},
  issue = {22},
  pages = {223401},
  numpages = {7},
  year = {2025},
  month = {Jun},
  publisher = {American Physical Society},
  doi = {10.1103/PhysRevLett.134.223401},
  url = {https://link.aps.org/doi/10.1103/PhysRevLett.134.223401}
}

@article{Guttridge2024,
  title = {Individual Assembly of Two-Species Rydberg Molecules Using Optical Tweezers},
  author = {Guttridge, Alexander and Hepworth, Tom R. and Ruttley, Daniel K. and Durst, Aileen A. T. and Eiles, Matthew T. and Cornish, Simon L.},
  journal = {Phys. Rev. Lett.},
  volume = {134},
  issue = {13},
  pages = {133401},
  numpages = {7},
  year = {2025},
  month = {Apr},
  publisher = {American Physical Society},
  doi = {10.1103/PhysRevLett.134.133401},
  url = {https://link.aps.org/doi/10.1103/PhysRevLett.134.133401}
}

@article{Domcke1993,
title = {Internal conversion funnel in benzene and pyrazine: adiabatic and diabatic representation},
journal = {Chem. Phys. Lett.},
volume = {203},
number = {2},
pages = {220-226},
year = {1993},
issn = {0009-2614},
doi = {https://doi.org/10.1016/0009-2614(93)85391-Z},
url = {https://www.sciencedirect.com/science/article/pii/000926149385391Z},
author = {W. Domcke and A.L. Sobolewski and C. Woywod},
}

@article{Koppel1983,
title = {Ultrafast non-radiative decay via conical intersections of molecular potential-energy surfaces: C2H4+},
journal = {Chem. Phys.},
volume = {77},
number = {3},
pages = {359-375},
year = {1983},
issn = {0301-0104},
doi = {https://doi.org/10.1016/0301-0104(83)85091-5},
url = {https://www.sciencedirect.com/science/article/pii/0301010483850915},
author = {H. Köppel}
}

@Article{Marciniak2015,
author={Marciniak, A.
and Despr{\'e}, V.
and Barillot, T.
and Rouz{\'e}e, A.
and Galbraith, M. C. E.
and Klei, J.
and Yang, C.-H.
and Smeenk, C. T. L.
and Loriot, V.
and Reddy, S. Nagaprasad
and Tielens, A. G. G. M.
and Mahapatra, S.
and Kuleff, A. I.
and Vrakking, M. J. J.
and L{\'e}pine, F.},
title={XUV excitation followed by ultrafast non-adiabatic relaxation in PAH molecules as a femto-astrochemistry experiment},
journal={Nat. Commun.},
year={2015},
month={Aug},
day={13},
volume={6},
number={1},
pages={7909},
issn={2041-1723},
doi={10.1038/ncomms8909},
url={https://doi.org/10.1038/ncomms8909}
}

@article{Harris1963,
    author = {Harris, Robert A.},
    title = {Predissociation},
    journal = {J. Chem. Phys.},
    volume = {39},
    number = {4},
    pages = {978-987},
    year = {1963},
    month = {08},
    issn = {0021-9606},
    doi = {10.1063/1.1734401},
    url = {https://doi.org/10.1063/1.1734401}
}

@article{Rice1933,
    author = {Rice, O. K.},
    title = {Predissociation and the Crossing of Molecular Potential Energy Curves},
    journal = {J. Chem. Phys.},
    volume = {1},
    number = {6},
    pages = {375-389},
    year = {1933},
    month = {06},
    issn = {0021-9606},
    doi = {10.1063/1.1749305},
    url = {https://doi.org/10.1063/1.1749305}
}

@book{Koonin1990,
  title={Computational Physics: Fortran Version},
  author={Steven E. Koonin and Dawn C. Meredith and William H. Press},
  year={1990},
  url={https://api.semanticscholar.org/CorpusID:121309039}
}

@book{thomas2013numerical,
  title={Numerical Partial Differential Equations: Finite Difference Methods},
  author={Thomas, J.W.},
  isbn={9781489972781},
  series={Texts in Applied Mathematics},
  url={https://books.google.de/books?id=83v1BwAAQBAJ},
  year={2013},
  publisher={Springer New York}
}

\end{document}